%% file: acl_latex.tex
\documentclass[11pt]{article}
\usepackage[preprint]{acl}
\usepackage{times}
\usepackage{pifont}
\usepackage{fontawesome5}
\usepackage[T1]{fontenc}
\usepackage[utf8]{inputenc}
\usepackage{microtype}
\usepackage{inconsolata}
\usepackage{amsmath,amssymb}
\DeclareFontShape{OMX}{cmex}{m}{n}{<->cmex10}{}
\usepackage{algorithm}
\usepackage{algpseudocode}
\usepackage{graphicx}
\usepackage{placeins}
\usepackage{booktabs}
\usepackage{tabularx}
\usepackage{multirow}
\usepackage{xcolor}
\usepackage{colortbl}
\usepackage[most]{tcolorbox}
\tcbuselibrary{skins}
\usepackage{enumitem}
\definecolor{headerblue}{RGB}{220, 240, 250}
\definecolor{headerpurple}{RGB}{235, 230, 250}

\renewcommand{\topfraction}{0.85}
\renewcommand{\dbltopfraction}{0.85}
\renewcommand{\floatpagefraction}{0.85}
\definecolor{promptGreen}{RGB}{120, 195, 120}
\definecolor{promptBlue}{RGB}{110, 120, 190}
\newtcolorbox{mypromptbox}[2][]{
    enhanced,
    drop shadow,
    colframe=#2,
    colback=white,
    coltitle=white,
    title=\textbf{\normalsize #1},
    fontupper=\small\ttfamily,
    sharp corners,
    boxrule=0.8pt,
    left=2mm, right=2mm, top=1.5mm, bottom=1.5mm,
    before skip=0pt, after skip=0pt,
    arc=0mm,
    outer arc=0mm
}

\makeatletter
\newcommand{\appendixfloatlayout}{%
\setlength{\@fptop}{0pt}\setlength{\@dblfptop}{0pt}%
\setlength{\@fpsep}{8pt}\setlength{\@dblfpsep}{8pt}%
\setlength{\textfloatsep}{10pt plus 2pt minus 2pt}%
\setlength{\floatsep}{8pt plus 2pt minus 2pt}%
\setlength{\intextsep}{8pt plus 2pt minus 2pt}%
\setlength{\abovecaptionskip}{4pt}\setlength{\belowcaptionskip}{0pt}%
\renewcommand{\topfraction}{0.95}\renewcommand{\bottomfraction}{0.9}%
\renewcommand{\textfraction}{0.05}\renewcommand{\floatpagefraction}{0.85}%
\renewcommand{\dbltopfraction}{0.95}\renewcommand{\dblfloatpagefraction}{0.85}%
\setcounter{topnumber}{4}\setcounter{bottomnumber}{3}\setcounter{totalnumber}{6}%
\setcounter{dbltopnumber}{3}}
\makeatother

\title{MAC-Bench: Benchmarking Multi-Agent Collaboration via\\ Interconnected Tasks}
\author{
  Zerui Chen\textsuperscript{1,\scalebox{1.5}{*}} \quad
  Qinggang Zhang\textsuperscript{3,\scalebox{1.5}{*}\faCube} \quad
  Qichuan Liu\textsuperscript{1} \quad
  Zhishang Xiang\textsuperscript{1} \quad
  Zhimin Wei\textsuperscript{1} \\
  \bfseries Yunbo Tang\textsuperscript{1} \quad
  Zhuo Li\textsuperscript{2} \quad
  Jinsong Su\textsuperscript{1} \quad
  Zhihong Zhang\textsuperscript{1,\scalebox{1.25}{\textdagger}} \\
  \textnormal{\textsuperscript{1}Xiamen University \quad
  \textsuperscript{2}ByteDance \quad
  \textsuperscript{3}Jilin University} \\
  {\small\ttfamily chenzerui1@stu.xmu.edu.cn \quad
  qinggangzhang@jlu.edu.cn \quad zhihong@xmu.edu.cn} \\
  {\small\normalfont \scalebox{1.5}{*}~Equal contribution. \quad
  \faCube~Project Leader. \quad
  \textdagger~Corresponding author.} \\[0.5em]
  {\small\normalfont
  \href{https://github.com/rzchen777-wq/MAC_Bench}{\textcolor{black}{\faGithub\enspace\textbf{Code}\,\textperiodcentered\,GitHub}}
  \quad\textcolor{gray}{\textbar}\quad
  \href{https://huggingface.co/datasets/sihuichuibuqi/MAC-Bench}{\textcolor{black}{\raisebox{-0.15em}{\includegraphics[height=1.15em]{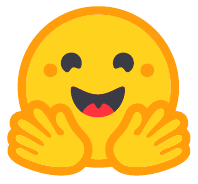}}\enspace\textbf{Dataset}\,\textperiodcentered\,Hugging Face}}}
}

\begin{document}
\maketitle
\begin{abstract}
\input{content/0-abstract}
\end{abstract}
\section{Introduction}
\label{sec:intro}
\input{content/1-introduction}
\input{content/2-preliminary}
\input{content/3-method}
\input{content/4-experiment}
\section{Conclusion}
\input{content/6-conclusion}
\bibliography{custom}
\clearpage
\appendix
\input{content/7-appendix}

\end{document}

%% file: content/0-abstract.tex
Large language model (LLM)-based multi-agent systems (MAS) tackle challenging tasks by dividing work among agents and coordinating their contributions toward shared objectives. Evaluating these systems therefore requires benchmarks that assess not only task-solving performance but also the ability to handle interdependent work. However, existing benchmarks often emphasize task difficulty without substantial collaboration requirements or evaluate only a limited range of predefined collaboration patterns. In this paper, we introduce \textbf{MAC-Bench}, a comprehensive benchmark for evaluating multi-agent collaboration through interconnected tasks. Rather than imposing predefined agent roles or communication protocols, MAC-Bench embeds collaboration requirements directly into task dependencies and shared-resource constraints. Specifically, we mine reusable collaboration structures from selected benchmark tasks at two levels. Local collaboration patterns capture recurring interactions among subtasks, while global task structures characterize how these interactions contribute to complete workflows. Through coarse-to-fine composition, these structures are recombined into task graphs combining different collaboration requirements. Tasks are instantiated in executable environments with automated verification of partial progress and complete task resolution.




%% file: content/1-introduction.tex
Large language model (LLM)-based multi-agent systems (MAS) combine task decomposition, specialized roles, and inter-agent communication to extend the capabilities of individual agents \citep{li2023camel,wu2023autogen,hong2023metagpt,qian2024chatdev,feng2026graph}. Recent systems further automate team organization and adaptation, achieving strong performance in reasoning, software engineering, and tool use \citep{ruan2026aorchestra,yang2025agentnet,hao2026metateam,cao2026evoteam}. 
The rapid advancement of LLM-based MAS calls for rigorous benchmarks to measure collaborative progress and guide system design on complex, long-horizon tasks.

\begin{figure*}[t]
\centering
\includegraphics[width=\textwidth]{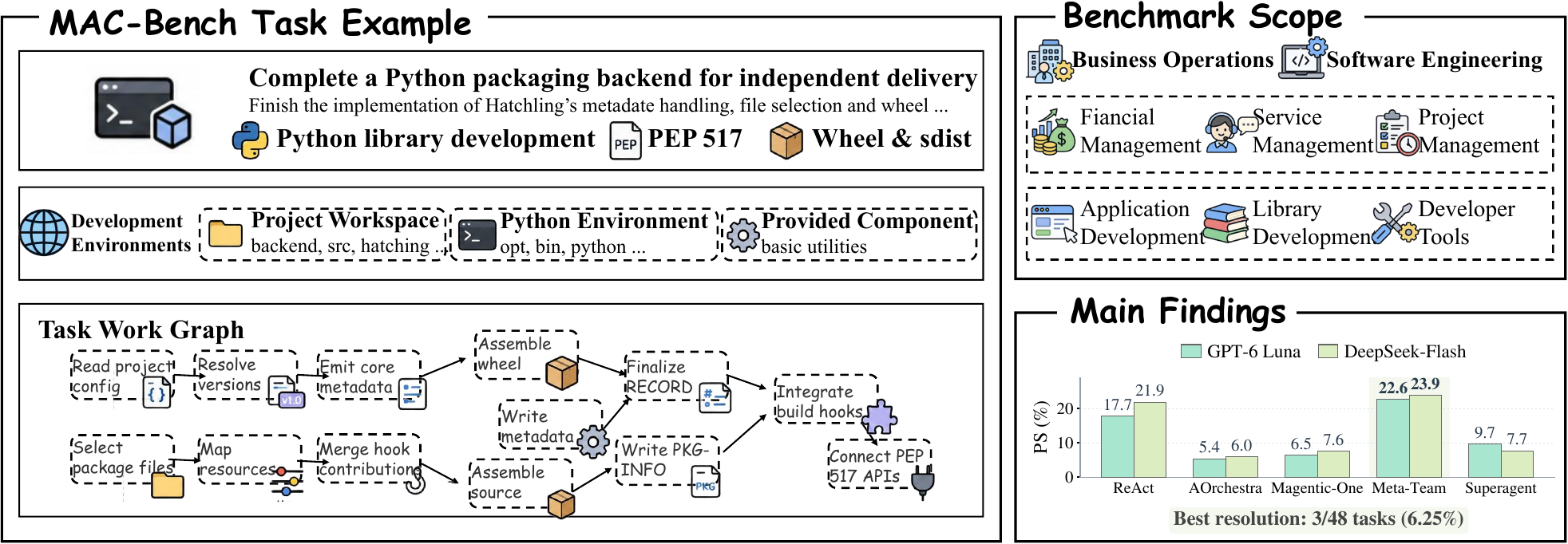}
\caption{\textbf{MAC-Bench overview.} Left: a packaging task links work branches through intermediate artifacts. Upper right: six task categories across two domains. Lower right: overall Progress Score (PS) for five systems with two LLM backbones; the highest Resolved Rate (RR) is 6.25\%.}
\vspace{-2mm}
\label{fig:intro}
\end{figure*}

Benchmarks used to evaluate MAS fall into two categories:
(1) \textit{General-purpose agent benchmarks} offer complex, multi-step tasks for evaluating MAS \citep{ruan2026aorchestra,yang2025agentnet,hao2026metateam}, spanning reasoning \citep{hendrycks2021math}, software engineering \citep{jimenez2024swebench}, and interactive tool use \citep{mialon2024gaia,trivedi2024appworld,li2023apibank,yao2024taubench,merrill2026terminalbench,zhou2024webarena,xie2024osworld}. Their difficulty and execution length motivate division of work to improve overall task performance. (2) \textit{MAS-specific benchmarks} construct explicit collaboration requirements \citep{zhu2025multiagentbench,grotschla2025agentsnet,du2026protocolbench,ke2026masorchestra}, including distributed-information and privacy-constrained exchange \citep{zhang2026silobench,ossowski2024comma,park2026pacbench}, role-separated execution \citep{kim2026teambench}, joint action in games \citep{xu2024magic,agashe2025llmcoordination,sun2025collabovercooked}, and integration of separately developed features \citep{khatua2026cooperbench}. Recent efforts extend evaluation to long-horizon workflows and scalable task generation \citep{kim2026teambench,shu2026agentworld,tessera2026alem}. Tasks are expanded through variants of curated templates \citep{kim2026teambench,shu2026agentworld}, or by generating coordination requirements from environment-specific mechanisms \citep{tessera2026alem}.

Despite this progress, constructing tasks for MAS collaboration evaluation presents two dilemmas. (1) Complexity does not imply interdependence \textbf{(Dilemma 1)}. Difficulty and execution length can increase through serial work without creating substantial branches that require coordination \citep{kim2025agentscaling,anthropic2025multiagentresearch}. (2) Limited patterns constrain collaboration coverage \textbf{(Dilemma 2)}. Testing selected interactions provides targeted evidence \citep{khatua2026cooperbench,kim2026teambench,shu2026agentworld,tessera2026alem}, but does not fully capture the complexity of real-world collaboration. This raises a central question: How can we construct executable tasks that test diverse combinations of collaboration requirements?

To overcome these fundamental dilemmas, we propose \textbf{MAC-Bench}, a benchmark for evaluating multi-agent collaboration through interconnected tasks. We construct its tasks by mining reusable structures from selected benchmark tasks \citep{shi2026saasbench,ren2026saasbench,xu2026roadmapbench,deng2026swemilestone} and recombining them to support compositional scaling of task structures. The extracted structures form a library at two scales: \ding{182}~\textbf{local collaboration patterns} that capture interactions among subtasks, and \ding{183}~\textbf{global task structures} that retain task organization and interface constraints. Coarse-to-fine composition recombines global structures and expands them with local patterns to form task graphs combining different collaboration requirements. The graphs guide instantiation in executable environments, where task outcomes can be verified.

%

Our contributions are summarized as follows:
\begin{itemize}
\item We introduce MAC-Bench, covering six categories in business operations and software engineering through complete, executable tasks with substantial, interdependent branches.
\item We develop a compositional construction approach that mines reusable local collaboration patterns and global task structures from existing tasks and recombines them into new combinations of collaboration requirements.
\item We evaluate four MAS and a single-agent baseline across two LLM backbones. Multi-agent designs show mixed benefits in verified progress, and no configuration resolves more than 6.25\% of the 48 test tasks.
\end{itemize}

\begin{figure}[t]
\centering
\includegraphics[width=\columnwidth]{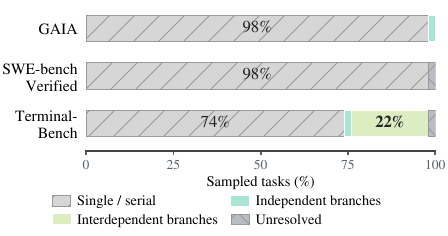}
\caption{\textbf{Task structures in general-purpose agent benchmarks.} Percentages over 50 sampled tasks per benchmark. Unresolved denotes uncertain classifications from available materials (Appendix~\ref{app:preliminary}).}
\label{fig:preliminary-structure}
\end{figure}

%% file: content/2-preliminary.tex
\section{Preliminary Study}
\label{sec:preliminary}

Before presenting MAC-Bench, we examine task structures for evaluating multi-agent collaboration and their coverage in existing benchmarks.


\subsection{What Tasks Test Collaboration?}
Evaluating multi-agent collaboration requires examining how agents coordinate different parts of a task toward a shared outcome \citep{hong2023metagpt,qian2024chatdev}. We describe these parts as branches, each involving sustained work toward a local goal and producing intermediate results. Separately progressable branches support division of work and parallel execution \citep{anthropic2025multiagentresearch,carlini2026parallelcompiler}. However, parallel progress alone does not establish the ability to coordinate interdependent work \citep{kim2025agentscaling}. When branches depend on one another's outputs or modify shared resources, local decisions can constrain work in other branches \citep{anthropic2026multiagentpatterns}. Agents must then reconcile their contributions, as misalignment and integration conflicts can prevent joint completion \citep{cemri2025mast,khatua2026cooperbench}. Tasks combining substantial branches with these dependencies allow evaluation of local progress and coordination toward task completion.


\begin{figure*}[t]
\centering
\includegraphics[width=\textwidth]{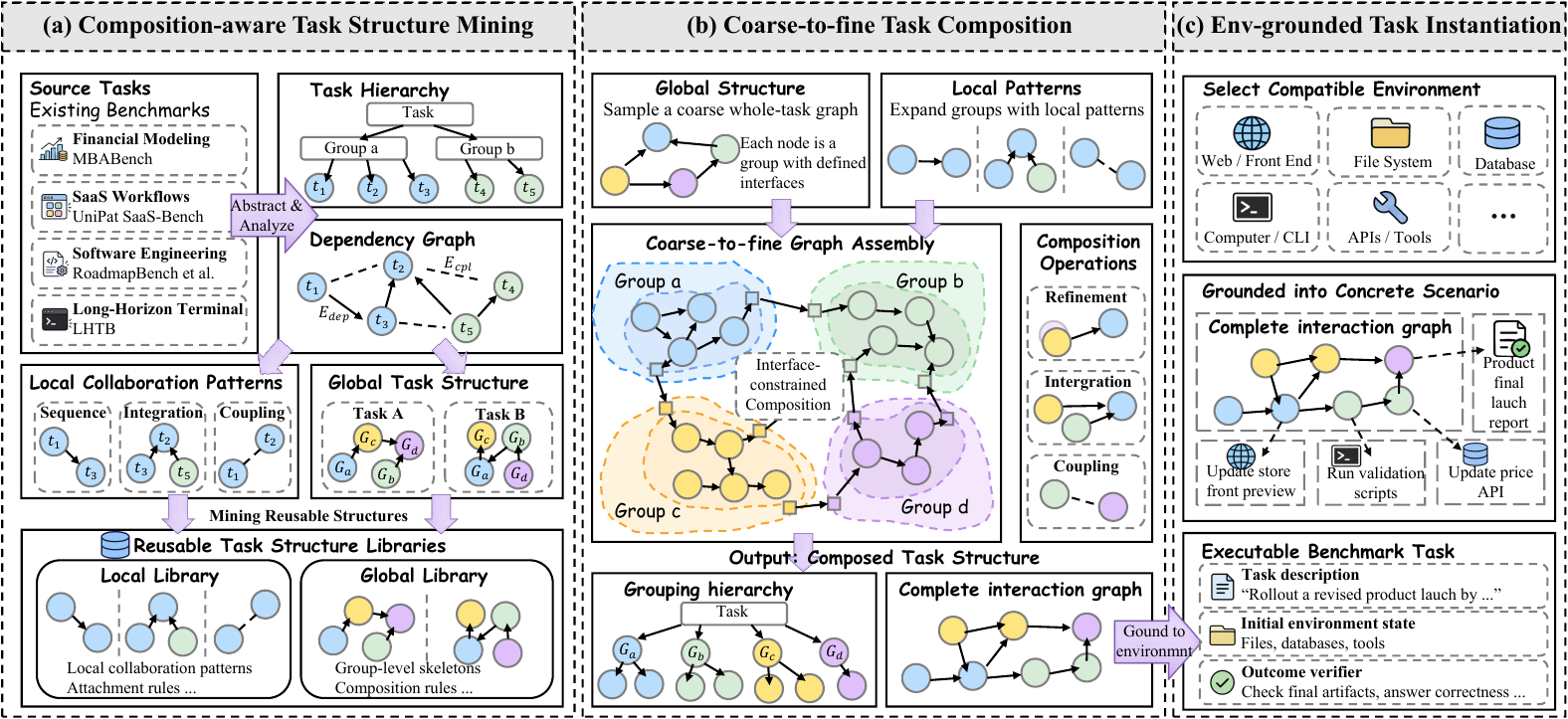}
\caption{\textbf{Construction pipeline of MAC-Bench.} (a) Mining extracts reusable local patterns and global task structures from existing tasks. (b) Composition recombines global structures and expands them with local patterns. (c) Instantiation specifies concrete tasks, initial environments, and outcome verifiers.}
\label{fig:method}
\end{figure*}

\subsection{\texorpdfstring{Do Existing Benchmarks\\Cover Such Tasks?}{Do Existing Benchmarks Cover Such Tasks?}}
We next examine how frequently benchmarks commonly used for MAS evaluation contain substantial branches with cross-branch dependencies. We analyze 50 uniformly sampled tasks from each of GAIA, SWE-bench Verified, and Terminal-Bench 2.0 \citep{mialon2024gaia,jimenez2024swebench,merrill2026terminalbench}, distinguishing serial work, independent branches, and interdependent branches (Appendix~\ref{app:preliminary}). As shown in Figure~\ref{fig:preliminary-structure}, single or primarily serial work accounts for 98\% of sampled tasks in both GAIA and SWE-bench Verified and 74\% in Terminal-Bench 2.0. Only 22\% of the Terminal-Bench samples have confirmed interdependent branches, motivating explicit construction of such tasks.

For MAS-specific benchmarks, we consider how different collaboration requirements are combined within complete tasks. Template-based expansion varies task content or initial conditions within established workflows \citep{kim2026teambench,shu2026agentworld}, while procedural environments generate coordination requirements through their own mechanics \citep{tessera2026alem}. Reusing a complete workflow preserves the connections among task branches, but carries over their original organization. An interaction alone, however, leaves unspecified the work that produces and consumes its result. Assessing such combinations calls for tasks that link local interactions to substantial work across branches.

These analyses motivate selecting tasks with substantial, interdependent branches as material for benchmark construction. Benchmarks of professional work and multi-component software development provide candidates for such selection \citep{shi2026saasbench,ren2026saasbench,xu2026roadmapbench,deng2026swemilestone}. We can then mine and recombine their constituent structures to create new combinations of collaboration requirements within coherent, executable tasks.

%% file: content/3-method.tex
%

\section{MAC-Bench}
\label{sec:method}

MAC-Bench evaluates multi-agent systems on executable tasks involving substantial and interdependent work. This section presents their format, evaluation, and construction.
\subsection{Task Format and Evaluation}
\label{sec:task-design}

As illustrated in Figure~\ref{fig:intro}, each MAC-Bench task involves multiple interconnected work components toward a common objective. An evaluation instance provides a task description, an initialized environment, and permitted operations. The evaluated system determines how to organize and coordinate the work, without prescribed agent roles or workflows. Construction graphs and reference materials remain hidden.

MAC-Bench contains 48 test tasks across six categories in business operations and software engineering, alongside a separate development set. Private outcome verifiers assess partial progress and complete resolution through observable environment states and artifacts. We construct these tasks through a three-stage pipeline that combines diverse work dependencies.

\subsection{Benchmark Construction}
\label{sec:benchmark-construction}

The construction process consists of three stages: (1) \textbf{Dependency-aware Task Structure Mining} abstracts tasks from existing benchmarks into graphs and builds a reusable library of local collaboration patterns, global task structures, and composition rules. (2) \textbf{Coarse-to-fine Task Composition} uses this library to recombine global structures and expand them with local patterns into new task graphs. (3) \textbf{Environment-grounded Task Instantiation} refines the composed graphs into environment-supported scenarios and prepares task descriptions, initial environment states, and outcome verifiers. The task graphs specify work requirements and dependencies; agent organization is left to the evaluated system. Figure~\ref{fig:method} illustrates the construction process.

\subsubsection{Dependency-aware Task Structure Mining}
\label{sec:induction}

Cross-task recombination requires reusable local interactions and the global structures and interface conditions that connect them into complete tasks. We abstract selected benchmark tasks \citep{shi2026saasbench,ren2026saasbench,xu2026roadmapbench,deng2026swemilestone} into graphs and extract local and global structures for composition.

\paragraph{Task Graph Abstraction.}

We first recursively decompose each task into subtasks with identifiable local outcomes, following hierarchical task decomposition \citep{erol1994umcp}. Given a task $\mathcal T$, each subtask is represented as $t=\langle g_t,i_t,o_t,a_t\rangle$, specifying its objective, required inputs, produced outputs, and read/write accesses to shared resources. A decomposition policy $\pi_{\mathrm{dec}}$ generates children of subtask $t$ conditioned on $t$ and the original task $\mathcal T$:
\begin{equation}
\{t_1,\ldots,t_m\}\sim\pi_{\mathrm{dec}}\!\left(\,\cdot\mid t,\mathcal T\right).
\label{eq:task-decomposition}
\end{equation}
Recursive application yields a task tree $\mathcal H=(V^H,E^H)$, where $V^H$ contains the original task and its subtasks, and $E^H$ contains parent--child decomposition edges. Its leaf subtasks form $\mathcal U\subseteq V^H$. Each non-root tree node $h$ identifies the group of leaf subtasks in its subtree.

We then infer relations among leaf subtasks from their inputs, outputs, and resource accesses. A dependency $t\prec t'$ indicates that $t$ supplies an output or establishes a resource condition required by $t'$. A coupling $t\leftrightarrow t'$ indicates mutually affecting modifications to shared resources. These relations define the annotated dependency graph:
\begin{equation}
\begin{aligned}
E^{\mathrm{dep}}&=\{(t,t')\mid t,t'\in \mathcal U,\ t\ne t',\ t\prec t'\},\\
E^{\mathrm{cpl}}&=\{\{t,t'\}\mid t,t'\in \mathcal U,\ t\ne t',\ t\leftrightarrow t'\},\\
\mathcal G&=(\mathcal U,E^{\mathrm{dep}},E^{\mathrm{cpl}}).
\end{aligned}
\label{eq:work-graph}
\end{equation}
Dependencies form a DAG; couplings are undirected annotations that do not prescribe an execution order. Thus, $\mathcal H$ records task grouping and $\mathcal G$ records subtask interactions. Appendix~\ref{app:structure-learning} details decomposition and abstraction checks.

\paragraph{Local Collaboration Pattern Mining.}

We next mine local collaboration patterns and rules for connecting them. To match patterns across tasks, we cluster leaf subtasks using weighted $k$-medoids over semantic, interface, and structural features. Each cluster defines a subtask type $\tau$; its observed input/output types and resource-access modes are retained for subsequent interface matching.

We enumerate connected two- and three-node subgraphs $\mathcal G_{\mathrm{sub}}$, retaining their subtask types, direct dependencies, coupling annotations, and precedence inherited from $\mathcal G$. Each receives a canonical representation $M$:
\begin{equation}
M=\min_{\sigma\in\mathcal P_k}e_\sigma(\mathcal G_{\mathrm{sub}}),
\label{eq:pattern-canonicalization}
\end{equation}
where $\mathcal P_k$ contains all permutations of the $k$ nodes, $e_\sigma$ encodes type labels $\tau$, direct and indirect precedence, and couplings in order $\sigma$, and the minimum is lexicographic. We merge identical representations into pattern types and filter by distinct-task support.

Within each task, we record overlapping occurrences of patterns $M$ and $M'$. Their node alignment $\psi$ maps shared subtasks between the two patterns' positions. Each tuple $(M,M',\psi)$ specifies a way to connect the patterns by merging their shared nodes. We retain these composition rules with their distinct-task support counts.

\paragraph{Global Task Structure Extraction.}

Finally, we extract global task structures to retain the context in which local patterns contribute to complete tasks. The root's children in $\mathcal H$ partition the subtasks of $\mathcal G$ into groups $h$. Contracting each group into one node yields a coarse graph $\overline{\mathcal G}$ with cross-group dependencies and coupling annotations. Each node retains its group's objective and boundary interfaces: external inputs and outputs, and read/write accesses to shared resources. The tree records containment; the coarse graph records interactions between groups.

Coarsening at non-root internal nodes yields refinement rules, each recording child groups, their relations, and parent-to-child interface mappings. Subtasks combining multiple upstream outputs provide integration templates, while shared-resource interactions provide coupling templates. These templates retain the relevant input/output types and access modes.

To link local patterns to these structures, we categorize groups $h$ by their objectives and boundary interfaces, assigning each a label $r$. Each pattern occurrence is indexed under the smallest categorized group containing its subtasks. For category $r$, $\mathcal M_r$ collects the indexed pattern types and $c_r(M)$ counts its distinct supporting tasks.

Together, these results form $\mathcal K=(\mathcal L_L,\mathcal L_G)$. The local library $\mathcal L_L$ stores subtask types and their interface characteristics, category-indexed patterns, and overlap-based composition rules with support statistics. The global library $\mathcal L_G$ stores coarse task structures, boundary interfaces, and refinement, integration, and coupling rules. Shared group categories link the two libraries for subsequent composition. Mining and indexing details are provided in Appendix~\ref{app:structure-learning}.

\subsubsection{Coarse-to-fine Task Composition}
\label{sec:synthesis}

Local patterns provide reusable interactions, while global structures specify how they are organized within complete tasks. Using $\mathcal K$, we first compose group-level structures from $\mathcal L_G$, expand each group with patterns from $\mathcal L_L$, and assemble local graphs using coarse dependencies and couplings.

\paragraph{Global Structure Composition.}

We sample a whole-task coarse graph from $\mathcal L_G$ with a corresponding grouping tree. Refinement replaces a group with its child-group graph through stored boundary mappings; integration inserts a multi-input group; coupling introduces shared-resource interactions. The tree records containment separately from graph interactions.

Target-guided search applies compatible rules to meet group and integration count ranges while preserving interfaces and dependency acyclicity. Accepted outputs comprise a grouping tree $\mathcal H^\star$ and a coarse graph $\overline{\mathcal G}^\star$, whose nodes correspond to terminal tree groups and retain category $r$, local size targets, and boundary interfaces.

\paragraph{Local Pattern Composition.}

We next expand each group node $h$ of $\overline{\mathcal G}^\star$ into a local subtask graph. For its category $r$, the patterns $\mathcal M_r$ in $\mathcal L_L$ define the seed candidates. Their distinct-task support counts give the sampling distribution
\begin{equation}
p(M\mid r)=\frac{c_r(M)+\alpha}
{\sum_{M'\in\mathcal M_r}(c_r(M')+\alpha)},
\label{eq:motif-sampling}
\end{equation}
where $\alpha>0$ is a smoothing constant. We sample a size-compatible seed from this distribution and expand it to cover the group's boundary requirements.

We expand the seed using connecting rules $(M,M',\psi)$ sampled by distinct-task support. Repeated attachments form a local graph $\mathcal G_h$ within the group's target size, preserving subtask types and acyclicity while satisfying requirements for its category and boundary interfaces.

\paragraph{Interface-constrained Graph Assembly.}

We map group boundaries to compatible nodes in $\mathcal G_h$ using interface characteristics in $\mathcal L_L$. These mappings realize coarse dependencies and couplings in the assembled graph $\widetilde{\mathcal G}$, retaining internal relations and group membership. After interface, resource-consistency, and acyclicity checks, $(\mathcal H^\star,\widetilde{\mathcal G})$ guides environment instantiation. Search, attachment, and interface-assignment details appear in Appendix~\ref{app:task-composition}.

\begin{table*}[t]
\centering
\small
\setlength{\tabcolsep}{3.2pt}
\renewcommand{\arraystretch}{1.02}
\resizebox{\textwidth}{!}{%
\begin{tabular}{l|cccccc|cccccc|cc}
\toprule
& \multicolumn{6}{c|}{\cellcolor{headerblue}\textbf{Business Operations}}
& \multicolumn{6}{c|}{\cellcolor{headerpurple}\textbf{Software Engineering}}
& \multicolumn{2}{c}{\textbf{Overall}} \\
\cmidrule(lr){2-7}\cmidrule(lr){8-13}\cmidrule(lr){14-15}
\multirow{2}{*}{\textbf{System}}
& \multicolumn{2}{c}{\textbf{Finance}}
& \multicolumn{2}{c}{\textbf{Service}}
& \multicolumn{2}{c|}{\textbf{Project}}
& \multicolumn{2}{c}{\textbf{App Dev.}}
& \multicolumn{2}{c}{\textbf{Lib Dev.}}
& \multicolumn{2}{c|}{\textbf{Dev. Tools}}
& \multicolumn{2}{c}{\textbf{All}} \\
& \textbf{PS}$\uparrow$ & \textbf{RR}$\uparrow$ & \textbf{PS}$\uparrow$ & \textbf{RR}$\uparrow$ & \textbf{PS}$\uparrow$ & \textbf{RR}$\uparrow$
& \textbf{PS}$\uparrow$ & \textbf{RR}$\uparrow$ & \textbf{PS}$\uparrow$ & \textbf{RR}$\uparrow$ & \textbf{PS}$\uparrow$ & \textbf{RR}$\uparrow$
& \textbf{PS}$\uparrow$ & \textbf{RR}$\uparrow$ \\
\midrule
\rowcolor{black!12}\multicolumn{15}{l}{\raisebox{-0.18\height}{\includegraphics[height=1.15em]{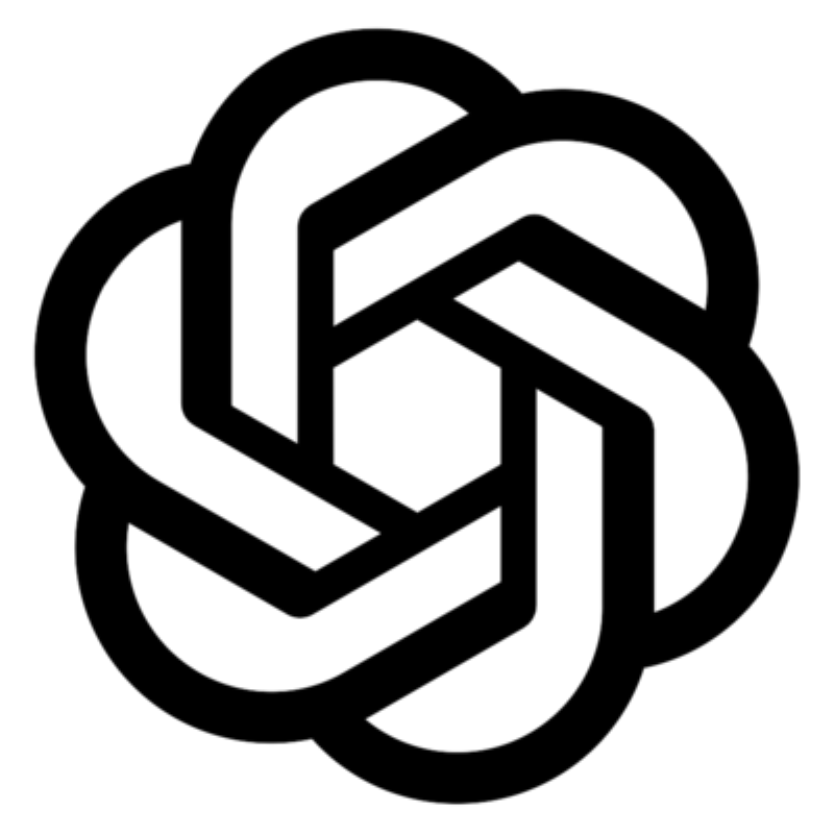}}\hspace{0.35em}\textbf{GPT-6 Luna}} \\
Single Agent (ReAct)~\citep{yao2023react} & \textbf{21.08} & 0.00 & \underline{24.93} & 0.00 & \underline{26.08} & \textbf{12.50} & \underline{7.79} & 0.00 & 8.08 & 0.00 & 18.50 & \textbf{12.50} & \underline{17.74} & \underline{4.17} \\
AOrchestra~\citep{ruan2026aorchestra} & 0.00 & 0.00 & 14.86 & 0.00 & 4.71 & 0.00 & \textbf{8.08} & 0.00 & 2.46 & 0.00 & 2.08 & 0.00 & 5.37 & 0.00 \\
Magentic-One~\citep{fourney2024magenticone} & 0.00 & 0.00 & 1.12 & 0.00 & 15.69 & 0.00 & 0.00 & 0.00 & 4.00 & 0.00 & 18.20 & 0.00 & 6.50 & 0.00 \\
Meta-Team~\citep{hao2026metateam} & \underline{12.14} & 0.00 & \textbf{33.19} & \textbf{12.50} & \textbf{42.69} & \textbf{12.50} & 7.08 & 0.00 & \textbf{16.20} & 0.00 & \underline{24.50} & \textbf{12.50} & \textbf{22.63} & \textbf{6.25} \\
Superagent~\citep{deepseek2026superagent} & 6.25 & 0.00 & 5.21 & 0.00 & 9.03 & 0.00 & 2.50 & 0.00 & \underline{10.00} & 0.00 & \textbf{25.00} & \textbf{12.50} & 9.66 & 2.08 \\
\rowcolor{black!12}\multicolumn{15}{l}{\raisebox{-0.18\height}{\includegraphics[height=1.15em]{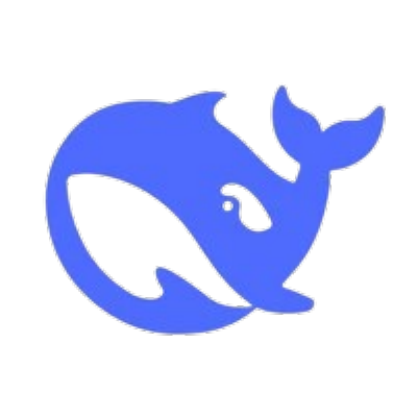}}\hspace{0.35em}\textbf{DeepSeek-Flash}} \\
Single Agent (ReAct)~\citep{yao2023react} & \textbf{30.71} & 0.00 & \underline{27.08} & \textbf{12.50} & \underline{32.08} & \textbf{12.50} & \textbf{8.93} & 0.00 & \underline{10.00} & 0.00 & \underline{22.50} & \textbf{12.50} & \underline{21.88} & \textbf{6.25} \\
AOrchestra~\citep{ruan2026aorchestra} & 0.00 & 0.00 & 17.71 & 0.00 & 4.86 & 0.00 & \underline{8.12} & 0.00 & 2.50 & 0.00 & 2.50 & 0.00 & 5.95 & 0.00 \\
Magentic-One~\citep{fourney2024magenticone} & 0.00 & 0.00 & 1.79 & 0.00 & 18.75 & 0.00 & 0.00 & 0.00 & 5.00 & 0.00 & 20.00 & 0.00 & 7.59 & 0.00 \\
Meta-Team~\citep{hao2026metateam} & 8.26 & 0.00 & \textbf{37.20} & 0.00 & \textbf{45.69} & \textbf{12.50} & 7.14 & 0.00 & \textbf{17.50} & 0.00 & \textbf{27.50} & \textbf{12.50} & \textbf{23.88} & \underline{4.17} \\
Superagent~\citep{deepseek2026superagent} & \underline{23.04} & 0.00 & 0.00 & 0.00 & 15.77 & 0.00 & 0.00 & 0.00 & 0.00 & 0.00 & 7.50 & 0.00 & 7.72 & 0.00 \\
\bottomrule
\end{tabular}%
}
\caption{\textbf{Main results on MAC-Bench.} Progress Score (PS) and Resolved Rate (RR) are percentages ($\uparrow$). Bold and underlined values indicate the best and second-best scores in each column within each backbone. Zero scores are left unmarked.}
\label{tab:main-results}
\end{table*}

\subsubsection{Environment-grounded Task Instantiation}
\label{sec:realization}

To construct executable tasks, we refine the composed task graphs into concrete scenarios in an environment $\Omega$ supporting the required resources and operations. Refinement specifies objectives, entities, and artifacts, adapting the work to the environment. The process produces:
\begin{equation}
(d,x_0,V)=\operatorname{Ground}_{\Omega}\!\left(\mathcal H^\star,\widetilde{\mathcal G}\right),
\label{eq:task-grounding}
\end{equation}
where $d$ describes the task objective, constraints, and requested outcomes; $x_0$ contains the initial records or project files; and $V$ checks task outcomes. Dependencies connect work through required intermediate artifacts or environment conditions, while shared-resource interactions impose consistency requirements across branches.

\paragraph{Outcome verification.}
Following state-based evaluation in executable agent environments \citep{trivedi2024appworld,tu2026scaleenv,aggarwal2026gymanything}, $V$ checks goal completion and consistency in final environment states and artifacts without prescribing an action sequence.

\paragraph{Task validation.}
We review task requirements and supported operations, check loading, reset, and initial non-completion, and calibrate outcome checks using valid, alternative, and invalid results. Appendices~\ref{app:task-instantiation} and~\ref{app:instantiation-diagnostics} report validation procedures and reference-execution records.

%% file: content/4-experiment.tex
\section{Experiments}
\label{sec:experiments}

We evaluate agent performance, analyze collaboration, and examine benchmark validity through three research questions: \textbf{RQ1 (Benchmark performance):} How do agent systems perform on MAC-Bench across task domains and backbones? \textbf{RQ2 (Collaboration analysis):} How do agent systems handle task dependencies and shared-resource constraints? \textbf{RQ3 (Benchmark validity):} Does MAC-Bench provide valid, diverse tasks for multi-agent collaboration evaluation?

\subsection{Experimental Setup}
\paragraph{Benchmark.}
The main evaluation uses all 48 test tasks in MAC-Bench, with eight tasks in each of six categories across Business Operations and Software Engineering (Section~\ref{sec:task-design}). Development tasks are separate from the test set.

\paragraph{Baselines.}
We compare single-agent ReAct \citep{yao2023react} with four MAS: AOrchestra \citep{ruan2026aorchestra}, Magentic-One \citep{fourney2024magenticone}, Meta-Team \citep{hao2026metateam}, and Superagent \citep{deepseek2026superagent}. Each system is evaluated using GPT-6 Luna and DeepSeek-Flash with thinking disabled; the same backbone is used for all agents and auxiliary model calls within a run.

\paragraph{Evaluation Protocol.}
We run one episode per system--backbone--task combination from a reset environment with identical instructions and tool access. Task caps of 5M, 10M, or 15M combined input and output tokens are preassigned by workload, cover all model calls, and apply to all systems and backbones. Each episode has a two-hour wall-clock limit. Meta-Team adapts on development tasks and freezes teams before testing. Further settings appear in Appendix~\ref{app:baselines}.

\paragraph{Evaluation Metrics.}
We report \textbf{Progress Score (PS)} and \textbf{Resolved Rate (RR)}. PS averages the fraction of achieved goals across tasks; RR is the fraction of fully resolved tasks. Automated checks assess final environment states and artifacts; full resolution also requires applicable preservation constraints. We report both as percentages by category and overall (Appendix~\ref{app:task-metrics}).

\begin{figure}[t]
\centering
\includegraphics[width=\columnwidth]{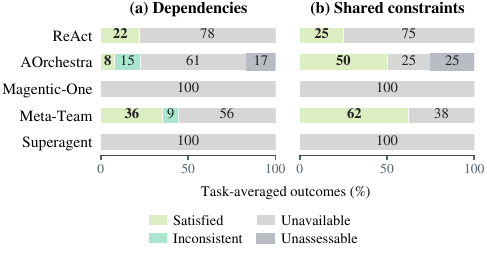}
\caption{\textbf{Task-relation outcomes with DeepSeek-Flash.} (a) 14 dependency checks on nine tasks. (b) Six shared-resource checks on four tasks. Percentages are averaged within tasks, then across tasks, and rounded. Unavailable means at least one required output is missing; unassessable denotes inconclusive evidence.}
\label{fig:relation-outcomes}
\end{figure}

\begin{figure}[t]
\centering
\includegraphics[width=\columnwidth]{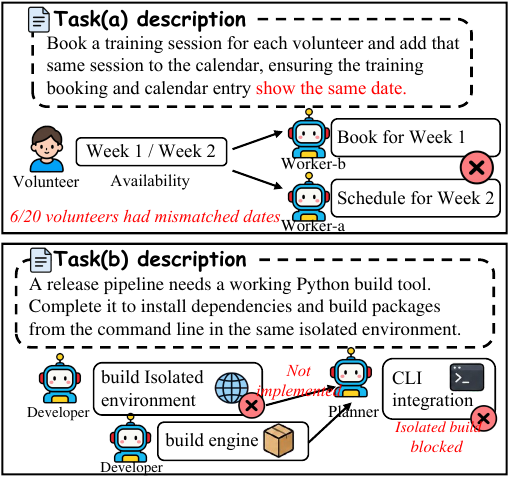}
\caption{\textbf{Meta-Team collaboration cases.} (a) Different workers leave booking and calendar dates inconsistent for six of 20 volunteers. (b) The planner's CLI calls the developer's unimplemented isolated-build environment constructor, blocking the isolated build.}
\label{fig:collaboration-cases}
\end{figure}

\begin{figure}[t]
\centering
\includegraphics[width=\columnwidth]{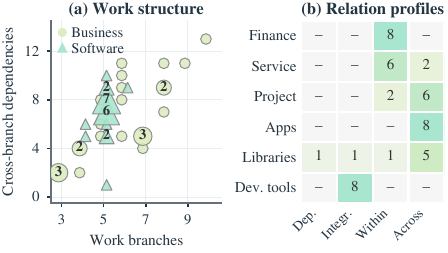}
\caption{\textbf{Work structures in MAC-Bench.} (a) Task frequencies by branch and cross-branch dependency counts. (b) Exclusive profiles: dependencies only (Dep.), integration without sharing (Integr.), or integration with sharing only within (Within) or across branches (Across). Dashes mark zero counts.}
\label{fig:final-task-structure}
\end{figure}

\subsection{Benchmark Performance (RQ1)}
\label{sec:main-results}
We compare single-agent and multi-agent systems across six task categories and two LLM backbones (Table~\ref{tab:main-results}). PS and RR reveal whether progress gains translate into complete resolution and how system rankings vary by category.

\paragraph{Obs. 1. Progress gains vary across multi-agent systems.}
Meta-Team leads in overall PS with GPT-6 Luna and DeepSeek-Flash, reaching 22.63\% and 23.88\%, respectively, versus ReAct's 17.74\% and 21.88\% (Table~\ref{tab:main-results}). AOrchestra, Magentic-One, and Superagent achieve lower overall PS than ReAct under both backbones.

\paragraph{Obs. 2. Progress and resolution yield different rankings.}
The highest RR is 6.25\%. With DeepSeek-Flash, ReAct resolves three tasks and Meta-Team two, despite Meta-Team's higher PS; this ordering reverses with GPT-6 Luna. Superagent resolves one task with GPT-6 Luna and none with DeepSeek-Flash; AOrchestra and Magentic-One resolve none under either backbone. Progress and complete resolution thus rank systems differently.

\paragraph{Obs. 3. Rankings vary across task categories.}
ReAct outperforms Meta-Team on Financial Management under both backbones, despite its lower overall PS. With GPT-6 Luna, Superagent ranks third overall but leads on Developer Tools with 25.00\% PS.

\subsection{Collaboration Analysis (RQ2)}
\label{sec:collaboration-analysis}
Beyond aggregate scores, we examine task dependencies and shared-resource constraints in five-system DeepSeek-Flash runs on nine business tasks (Appendix~\ref{app:graph-collaboration-analysis}). Figure~\ref{fig:relation-outcomes} distinguishes satisfied relations, inconsistent results, unavailable outputs, and unassessable cases, averaging within tasks before across tasks. Figure~\ref{fig:collaboration-cases} complements these statistics with business and software examples.

\paragraph{Obs. 4. Results can remain inconsistent across branches.}
Meta-Team's dependency outcomes are 35.6\% satisfied, 8.9\% inconsistent, and 55.6\% unavailable (Figure~\ref{fig:relation-outcomes}). In the market task, AOrchestra workers produce internally consistent allocations and inspections, yet 39 of 48 inspections disagree with their upstream allocations. Local consistency thus need not extend across branches.

\paragraph{Obs. 5. Cross-agent work can leave task dependencies unmet.}
In Figure~\ref{fig:collaboration-cases}(a), correction reduces date mismatches from eleven to six, but bookings and calendars still disagree for trainees accepting either week. In (b), the planner implements CLI integration while the developer leaves the required isolated-build environment constructor unimplemented, causing validation to fail.

\subsection{Benchmark Validity (RQ3)}
\label{sec:controlled-generation}
We validate MAC-Bench through task-structure analysis, reference execution, and scoring calibration. Across all 48 tasks, we examine branches, dependencies, and shared-resource constraints. Reference solutions pass complete task grading on all 48 tasks; none of the untouched initial states passes. For scoring calibration, we select two tasks per category and evaluate six conditions: a reference result, a legal alternative, missing work, a broken relation, a preservation violation, and a corresponding repair. These yield 72 evaluations; procedures and interventions are detailed in Appendix~\ref{app:instantiation-diagnostics}.

\paragraph{Obs. 6. Tasks combine several forms of interdependence.}
As shown in Figure~\ref{fig:final-task-structure}, tasks contain 3--10 substantial branches (median five). Multi-input integration appears in 47 tasks, and shared-resource or mutable-state constraints in 38: 21 across branches and 17 within branches only. Profiles vary across categories: Developer Tools combines dependencies with integration, whereas all eight Application Development tasks also require sharing resources across branches.

\begin{table}[t]
\centering\small
\renewcommand{\arraystretch}{1.12}
\setlength{\tabcolsep}{2pt}
\begin{tabular*}{\columnwidth}{@{\extracolsep{\fill}}lrrr@{}}
\toprule
\multirow{2}{*}{\textbf{Task}} & \multicolumn{3}{c}{\textbf{PS after violation (\%)}} \\
\cmidrule(lr){2-4}
& \shortstack{Missing\\work} & \shortstack{Broken\\relation} & \shortstack{Preservation\\violation} \\
\midrule
\rowcolor{black!5}\multicolumn{4}{@{}l}{\strut\textit{Business operations}} \\
Studio closeout & 85.7\phantom{\textsuperscript{$\dagger$}} & 85.7\phantom{\textsuperscript{$\dagger$}} & 100.0\textsuperscript{$\dagger$} \\
Workspace recovery & 85.7\phantom{\textsuperscript{$\dagger$}} & 85.7\phantom{\textsuperscript{$\dagger$}} & 100.0\textsuperscript{$\dagger$} \\
Housing allocation & 75.0\phantom{\textsuperscript{$\dagger$}} & 75.0\phantom{\textsuperscript{$\dagger$}} & 100.0\textsuperscript{$\dagger$} \\
Market allocation & 83.3\phantom{\textsuperscript{$\dagger$}} & 83.3\phantom{\textsuperscript{$\dagger$}} & 100.0\textsuperscript{$\dagger$} \\
Restaurant publication & 83.3\phantom{\textsuperscript{$\dagger$}} & 83.3\phantom{\textsuperscript{$\dagger$}} & 100.0\textsuperscript{$\dagger$} \\
Riverbank activities & 80.0\phantom{\textsuperscript{$\dagger$}} & 80.0\phantom{\textsuperscript{$\dagger$}} & 100.0\textsuperscript{$\dagger$} \\
\addlinespace[2pt]
\rowcolor{black!5}\multicolumn{4}{@{}l}{\strut\textit{Software engineering}} \\
Black & 20.0\phantom{\textsuperscript{$\dagger$}} & 80.0\phantom{\textsuperscript{$\dagger$}} & 40.0\phantom{\textsuperscript{$\dagger$}} \\
Build & 20.0\phantom{\textsuperscript{$\dagger$}} & 40.0\phantom{\textsuperscript{$\dagger$}} & 0.0\phantom{\textsuperscript{$\dagger$}} \\
Jinja & 0.0\phantom{\textsuperscript{$\dagger$}} & 60.0\phantom{\textsuperscript{$\dagger$}} & 100.0\textsuperscript{$\dagger$} \\
Lark & 0.0\phantom{\textsuperscript{$\dagger$}} & 80.0\phantom{\textsuperscript{$\dagger$}} & 0.0\phantom{\textsuperscript{$\dagger$}} \\
FreshRSS & 0.0\phantom{\textsuperscript{$\dagger$}} & 75.0\phantom{\textsuperscript{$\dagger$}} & 75.0\phantom{\textsuperscript{$\dagger$}} \\
Kanboard & 0.0\phantom{\textsuperscript{$\dagger$}} & 75.0\phantom{\textsuperscript{$\dagger$}} & 50.0\phantom{\textsuperscript{$\dagger$}} \\
\bottomrule
\end{tabular*}
\caption{\textbf{Scoring under controlled changes.} All displayed results fail complete grading; $\dagger$ denotes PS of 100\% despite a failed preservation check. References, legal alternatives, and repairs pass on all twelve tasks with PS of 100\%.}
\label{tab:verifier-calibration}
\label{tab:rq3-validation-coverage}
\end{table}

\paragraph{Obs. 7. Scoring reflects goal completion and preservation requirements.}
All twelve legal alternatives and twelve repairs pass, while all 36 erroneous results are rejected. Table~\ref{tab:verifier-calibration} shows the goal-level progress retained under each violation. In the housing test, changing an allocation without updating its handover fails the handover goal while the other three goals remain complete, yielding PS of 75\%. Correcting the handover restores PS to 100\% and full acceptance.

Preservation checks additionally assess whether completed work leaves protected records and existing behavior intact. In Jinja, altering existing error metadata leaves all five task goals satisfied but fails the preservation check: PS remains 100\%, while the task is rejected. The dagger-marked entries distinguish this outcome from missing goal progress. Further conflict-and-repair tests for shared resources appear in Appendix~\ref{app:task-relationships}.

\section{Related Work}
\label{sec:related-work}

\paragraph{LLM-based multi-agent systems.}
LLM-based MAS coordinate specialized agents through communication and delegation \citep{li2023camel,wu2023autogen,hong2023metagpt,qian2024chatdev,chen2026legalgraphrag,wu2026memgraphrag}. G-Designer generates task-adaptive communication topologies, AFlow searches executable workflows, and MaAS samples query-dependent agent architectures \citep{zhang2024gdesigner,zhang2024aflow,zhang2025maas}. AOrchestra and Meta-Team further explore dynamic sub-agent creation and team-level adaptation \citep{ruan2026aorchestra,hao2026metateam}. These approaches design agent systems; MAC-Bench specifies task requirements and leaves agent organization open.

\paragraph{Agent benchmarks.}
General-purpose benchmarks assess reasoning, software engineering, and tool use \citep{mialon2024gaia,jimenez2024swebench,trivedi2024appworld}. WorkArena++ and TheAgentCompany extend evaluation to compositional knowledge work and workplace activities \citep{boisvert2024workarenaplus,xu2024agentcompany}. MAS-specific benchmarks test communication, distributed information, and coordinated execution \citep{ossowski2024comma,zhang2026silobench,zhu2025multiagentbench,kim2026teambench,khatua2026cooperbench}. Recent environments extend collaboration to long-horizon tasks \citep{shu2026agentworld}.

\paragraph{Task and environment generation.}
AgentGen generates environments and planning tasks \citep{hu2024agentgen}. EnvScaler and ScaleEnv synthesize tool-interactive environments and validated tasks \citep{song2026envscaler,tu2026scaleenv}. SWE-smith generates coding tasks through repository modifications, while Gym-Anything converts software into agent environments \citep{yang2025swesmith,aggarwal2026gymanything}. WorkArena++ and $\tau^2$-bench compose atomic components into larger tasks \citep{boisvert2024workarenaplus,barres2025tau2bench}.

%% file: content/6-conclusion.tex
MAC-Bench provides substantial, interconnected tasks across business operations and software engineering by mining and composing reusable task structures. Tasks combine varied dependencies, integration, and shared-resource constraints; targeted checks show effects of selected relation violations. Across two LLM backbones, Meta-Team leads in overall PS, while AOrchestra, Magentic-One, and Superagent trail ReAct. No configuration resolves more than 6.25\% of tasks. MAC-Bench separates partial progress from complete task resolution.

%% file: content/7-appendix.tex
\appendixfloatlayout
\section{Frequently Asked Questions}
\label{app:faq}

\paragraph{How do task structures relate to agent organization?}
Task structures specify how work results depend on one another; the evaluated system decides how agents produce and coordinate those results. A system can assign several branches to one agent or divide a branch among multiple agents. It receives the public task description, initialized environment, and permitted operations, while construction graphs and reference materials remain hidden. Outcome verifiers assess the required states and artifacts. Relation-level analysis then connects these outcomes to the system's actual division of work (Section~\ref{sec:task-design} and Appendix~\ref{app:graph-collaboration-analysis}).

\paragraph{How are task feasibility and scoring validated?}
We review the correspondence between public requirements, available task materials, and scoring checks. Reference solutions pass complete grading on all 48 test tasks. On twelve tasks, controlled scoring tests accept valid alternative results and repairs while rejecting missing work, broken relations, and preservation violations. Further tests examine shared-resource conflicts and missing upstream records. Together, these checks assess whether tasks can be completed and whether scoring responds to the stated requirements (Table~\ref{tab:verifier-calibration} and Appendix~\ref{app:instantiation-diagnostics}).

\paragraph{Why are task-resolution rates low?}
MAC-Bench tasks are demanding in specification length, workload, and coordination. Lengthy task descriptions and supporting materials contain requirements, exceptions, and constraints that systems must interpret and track throughout execution. Each task contains 3--10 substantial work branches, with a median of five. For example, the housing task covers 48 households: each needs an eligible flat, a reconciled account, a staffed key-collection appointment, and a consistent handover pack. Changing a flat assignment can affect the fee, collection location, and officer assignment. Software tasks similarly require interacting components to work together; the build-frontend task connects its CLI, build engine, and isolated environment. Complete resolution requires all task goals and applicable preservation constraints to hold together, combining substantial local work with consistency across its results. Task structures and examples are detailed in Appendix~\ref{app:benchmark-statistics}.

\paragraph{Why use existing tasks to construct a new benchmark?}
Existing tasks provide concrete examples of professional work and multi-component software development. We extract their local collaboration patterns and global task structures, then recombine and instantiate them into new evaluation tasks. This allows recurring dependencies and resource interactions to appear in different task organizations. The mining and held-out corpora are separate from the 48-task evaluation collection. Source composition and structural reuse analyses appear in Appendices~\ref{app:source-statistics} and~\ref{app:additional-experiments}.

\section{Benchmark Description and Statistics}
\label{app:benchmark-statistics}
This section describes the final MAC-Bench task collection, its categories and work structures, and the validation protocol. Construction sources are reported separately from the evaluation tasks.

\subsection{Benchmark Overview}
\label{app:world-statistics}
\begin{figure}[tbp]
\centering
\includegraphics[width=.9\columnwidth]{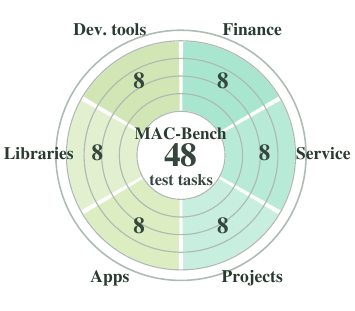}
\caption{\textbf{Task distribution in MAC-Bench.} Six equally sized categories contain eight test tasks each, totaling 48 tasks.}
\label{fig:dataset-branch-profile}
\end{figure}

The evaluation collection contains 48 test tasks, with eight tasks in each of six categories. The Business Operations domain contains 24 tasks across Financial Management, Service Management, and Project Management. The Software Engineering domain contains 24 tasks across Application Development, Library Development, and Developer Tools. Figure~\ref{fig:dataset-branch-profile} summarizes the category distribution. Development tasks are kept separate from the test set and used for method adaptation.

Software engineering environments expose a project workspace and terminal operations. Business operations environments expose native application records through application discovery, tool descriptions, calls, and result retrieval. Each accepted instance has a public task, an initial environment, and evaluator-side outcome checks. The task description states the required outcomes and available evidence; generated graph annotations and reference materials remain evaluator-side. Each task package contains the public description, environment assets, and outcome checks.

\subsection{Task Categories and Examples}
\label{app:task-categories}
The categories describe the work to be completed. Business tasks require changes to application records and consistent operational commitments. Software tasks require working implementations across interacting components. Table~\ref{tab:category-examples} gives representative requirements.

\begin{table}[!t]
\centering\small
\setlength{\tabcolsep}{4pt}
\renewcommand{\arraystretch}{1.13}
\begin{tabularx}{\columnwidth}{@{}>{\raggedright\arraybackslash}p{.32\columnwidth}>{\raggedright\arraybackslash}X@{}}
\toprule
\textbf{Category} & \textbf{Representative work and relations} \\
\midrule
\rowcolor{gray!10}\multicolumn{2}{@{}l}{\textbf{Business Operations}} \\
Financial\newline Management & Resolve credits, expenses, and balances; align recovery commitments with capacity and funding. \\
\addlinespace[3pt]
Service\newline Management & Reconcile eligibility, appointments, and allocations; keep bookings and schedules consistent. \\
\addlinespace[3pt]
Project\newline Management & Coordinate permissions, content, staffing, and costs; make delivery records reflect selected arrangements. \\
\midrule
\rowcolor{gray!10}\multicolumn{2}{@{}l}{\textbf{Software Engineering}} \\
Application\newline Development & Implement interacting data, service, and user-facing features while retaining existing behavior. \\
\addlinespace[3pt]
Library\newline Development & Complete parsing, validation, conversion, or rendering components supporting downstream execution. \\
\addlinespace[3pt]
Developer Tools & Implement formatting, model generation, or build components supporting command-line or release operations. \\
\bottomrule
\end{tabularx}
\caption{\textbf{Task categories in MAC-Bench.} Each category contains eight test tasks. Examples summarize public requirements from the final collection.}
\label{tab:category-examples}
\end{table}

For example, a workspace-settlement task combines invoice remedies, vendor settlement, expense ownership, and funded recovery commitments. A GraphQL task combines document processing, validation, input coercion, and request execution. In both cases, completing isolated components is insufficient when their results must agree or support downstream work. Concrete execution cases are analyzed in Section~\ref{sec:collaboration-analysis}.

\subsection{Task Structure Statistics}
\label{app:final-task-analysis}
We analyze the 48 test tasks using their public requirements and task contracts. LLM-assisted annotations identify work branches and their relations, retaining source quotations and contract entries for each task. We aggregate these annotations to obtain the counts below. Reviewing the revised release materials preserves the branch groupings and relation counts; the accompanying records link each annotation to its current public sources.

\paragraph{Branch and relation definitions.}
A substantial branch pursues a distinct local outcome. Serial identity and eligibility checks for the same service case are merged; pure final-handover aggregation is excluded from branch counts. Software branches follow the declared implementation responsibilities and interfaces. Directed relations represent required upstream artifacts or interface compatibility. Figure~\ref{fig:final-task-structure}(a) counts only dependencies between substantial branches; the full annotation also retains handover relations. Multi-input integration requires contributions from at least two upstream work units.

Shared constraints concern competing resource use or consequential updates to mutable state, rather than shared reads or unrelated records in one container. We distinguish constraints within a branch from those across branches. The four profiles in Figure~\ref{fig:final-task-structure}(b) are mutually exclusive: dependencies only; dependencies with integration and no sharing; integration with sharing confined to a branch; and integration with cross-branch sharing, possibly also within-branch constraints. All observed sharing profiles include integration.

\begin{table}[!t]
\centering\small
\setlength{\tabcolsep}{3pt}
\renewcommand{\arraystretch}{1.12}
\begin{tabularx}{\columnwidth}{@{}>{\raggedright\arraybackslash}Xrrrr@{}}
\toprule
{\bfseries \multirow{2}{*}{Category}} & {\bfseries \multirow{2}{*}{\shortstack{Branch\\range}}} & {\bfseries \multirow{2}{*}{Integr.}} & \multicolumn{2}{c}{\bfseries Sharing} \\
\cmidrule(lr){4-5}
 &  &  & {\bfseries Within} & {\bfseries Across} \\
\midrule
Finance & 7--10 & 8 & 8 & 0 \\
Service & 3--6 & 8 & 6 & 2 \\
Projects & 5--8 & 8 & 2 & 6 \\
Applications & 4--7 & 8 & 0 & 8 \\
Libraries & 5--5 & 7 & 1 & 5 \\
Dev. tools & 5--5 & 8 & 0 & 0 \\
\midrule
\rowcolor{gray!10}All tasks & 3--10 & 47 & 17 & 21 \\
\bottomrule
\end{tabularx}
\caption{\textbf{Final-task structure summary.} Each category contains eight tasks. Integr. counts tasks with multi-input integration. Share within counts tasks with within-branch constraints only; Share across counts tasks with cross-branch constraints.}
\label{tab:final-task-structures}
\end{table}

Table~\ref{tab:final-task-structures} summarizes the final collection. Median branch count is five. One library task has dependencies without multi-input integration; ten tasks have no shared constraint. Interface interpretation and reasonable alternative groupings can change fine-grained edge counts. The annotations retain the supporting task passages and grouping decisions.

\suppressfloats[t]
\begin{table}[!t]
\centering
\small
\renewcommand{\arraystretch}{1.12}
\begin{tabularx}{\columnwidth}{@{}>{\raggedright\arraybackslash}Xrr@{}}
\toprule
{\bfseries Source} & {\bfseries Mining} & {\bfseries Held-out} \\
\midrule
MBABench~\citep{yen2026mbabench} & 17 & 9 \\
UniPat SaaS-Bench~\citep{shi2026saasbench} & 12 & 5 \\
RoadmapBench~\citep{xu2026roadmapbench} & 30 & 15 \\
SaaSBench~\citep{ren2026saasbench} & 20 & 9 \\
SWE-Milestone~\citep{deng2026swemilestone} & 19 & 11 \\
LHTB~\citep{li2026lhtb} & 2 & 1 \\
\midrule
\rowcolor{gray!10}Total & 100 & 50 \\
\bottomrule
\end{tabularx}
\caption{Source-task composition of the mining and held-out corpora. Counts refer to accepted task graphs, rather than instantiated benchmark tasks.}
\label{tab:source-corpus}
\end{table}

\subsection{Validation Protocol}
\label{app:benchmark-validation}
\paragraph{Task and interface review.}
We check that public requirements identify the requested changes, preservation obligations, and evidence needed to complete the task. Cross-application identifiers, repository dependencies, and permitted operations are checked against the packaged environment. Each outcome is linked to the state fields or artifacts used to assess it.

\paragraph{Environment and scoring checks.}
Loading and reset checks verify the packaged starting state. Private graders run separately from the solver and assess exported application states or software artifacts. We evaluate both the untouched starting state and a completed reference result, then test scoring with valid alternatives and controlled violations. Constructed final states test the grader's decisions; reference executions test completion through the available environment. Appendix~\ref{app:instantiation-diagnostics} reports coverage, procedures, and results.
\label{app:human}

\subsection{Construction Sources}
\label{app:source-statistics}
Construction draws on professional-work and multi-component software tasks from six sources. These materials are separate from the 48-task evaluation collection. Public task specifications and accompanying documents, data, or code are converted into task hierarchies and annotated dependency graphs. Table~\ref{tab:source-corpus} gives the accepted mining and held-out composition. Tasks sharing a repository, original case, or normalized specification remain on the same side. The two partitions contain 65 and 34 source groups, respectively. The mining graphs have 4--38 leaves (median 9), and held-out graphs have 4--38 leaves (median 8). Median directed dependency counts are 12 and 10. These are subtask-level statistics, distinct from the terminal-group counts used in controlled generation.

The 100-task partition supplies the structure libraries; the 50-task partition is reserved for held-out reuse analysis. Extraction details and library sizes are reported in Appendices~\ref{app:structure-learning} and~\ref{app:controlled-generation-details}.

\setcounter{topnumber}{1}
\setcounter{totalnumber}{2}
\section{Additional System Evaluation}
\label{app:system-analysis}

The main evaluation measures completed goals and full task resolution. We first examine how these outcomes vary across tasks and how much execution they require. We then connect the dependency and shared-resource checks in RQ2 to agents' contributions, asking whether related results remain consistent when different agents produce them. Appendix~\ref{app:analysis-protocols} gives the analysis settings.

\subsection{Benchmark Performance (RQ1)}
\subsubsection{Task-level Performance Differences}
\label{app:paired-system-results}

\begin{figure}[tbp]
\centering
\includegraphics[width=\columnwidth]{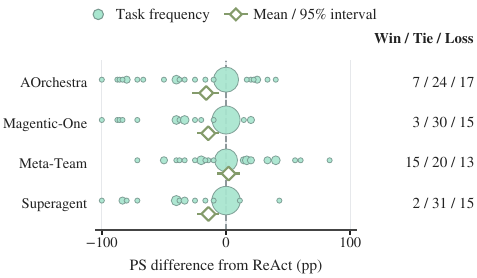}
\caption{\textbf{Task-level PS differences from ReAct.} Each system is compared on the same 48 tasks. Circle area indicates frequency; diamonds and lines show means and 95\% task-resampling intervals. Win, Tie, and Loss count tasks with higher, equal, and lower PS than ReAct.}
\label{fig:paired-system-progress}
\end{figure}

An overall PS difference can reflect broad improvements across tasks or gains on a few tasks offset by losses elsewhere. We therefore compare each MAS with ReAct on each of the 48 tasks. Figure~\ref{fig:paired-system-progress} plots their PS differences, with means and 95\% intervals from 20,000 task-bootstrap draws.

\paragraph{Obs. 8. Multi-agent gains vary across tasks.}
Meta-Team improves on 15 tasks, ties on 20, and performs worse on 13; its 2.00-point mean advantage over ReAct has a 95\% interval of $[-6.14, 10.05]$. AOrchestra, Magentic-One, and Superagent improve on seven, three, and two tasks, respectively. Superagent ties ReAct on 31 tasks and trails it on 15, with a mean difference of $-14.17$ points. Overall progress gains thus coexist with losses on individual tasks.

\subsubsection{Resource Use and Stopping Conditions}

The work branches in MAC-Bench require sustained execution. We report cumulative model usage and episode stopping conditions to characterize the resources used to pursue these outcomes.

\paragraph{Obs. 9. Most recorded episodes end at the shared token cap.}
Token-limit stopping accounts for 34--44 of 48 episodes per system (Table~\ref{tab:analysis-resources}). Superagent reaches the cap on 42 tasks; its median usage is 15.02M tokens and 132 model calls, the lowest call count among the systems. Fewer model calls therefore need not imply lower token use. The preassigned 5M, 10M, and 15M token caps are shared across systems on each task. An admitted final call may cross the cap. Episode stopping is recorded separately from the final task score.

\begin{table}[t]
\centering\small
\setlength{\tabcolsep}{2.6pt}
\renewcommand{\arraystretch}{1.12}
\begin{tabularx}{\columnwidth}{@{}>{\raggedright\arraybackslash}Xrrrrr@{}}
\toprule
{\bfseries \multirow{2}{*}{System}} & \multicolumn{2}{c}{\bfseries Median use} & \multicolumn{3}{c}{\bfseries Episode stops} \\
\cmidrule(lr){2-3}\cmidrule(lr){4-6}
 & {\bfseries Tokens (M)} & {\bfseries Calls} & {\bfseries Cap} & {\bfseries End} & {\bfseries Other} \\
\midrule
ReAct & 13.38 & 222 & 34 & 8 & 6 \\
AOrchestra & 15.00 & 995 & 44 & 3 & 1 \\
Magentic-One & 15.01 & 306 & 42 & 3 & 3 \\
Meta-Team & 14.49 & 297.5 & 35 & 13 & 0 \\
Superagent & 15.02 & 132 & 42 & 2 & 4 \\
\bottomrule
\end{tabularx}
\caption{Resource use and stopping across the 48 tasks. Tokens and Calls are median total tokens and recorded model calls. Cap, End, and Other count token-limit stops, normal episode endings, and other stops (including blocked and failed); each row totals 48. End does not imply task resolution.}
\label{tab:analysis-resources}
\end{table}

\subsection{Collaboration Analysis (RQ2)}
\label{app:contributor-analysis}

Task dependencies connect the results produced by different work branches. We examine where these connections remain unfinished, whether available results agree, and how their agreement changes during execution. The analysis retains all fourteen dependency checks across the same nine tasks and five systems used in RQ2. Each check covers the relevant people, activities, or allocations. Appendix~\ref{app:graph-collaboration-analysis} describes record matching and contributor attribution.

\begin{figure}[t]
\centering
\includegraphics[width=\columnwidth]{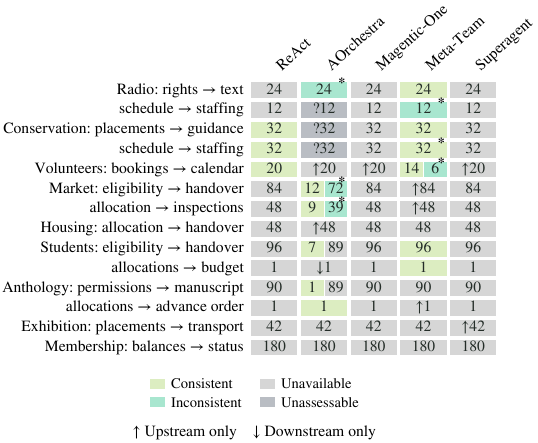}
\caption{\textbf{Dependent results by task relation.} Cells give item counts in each state; split cells show multiple states. For unavailable results, $\uparrow$ and $\downarrow$ indicate that only the upstream or downstream side is available; unmarked gray cells indicate neither. Stars mark relations with complete paired outputs and distinct agent contributors, traced in Table~\ref{tab:graph-contributions}.}
\label{fig:dependency-detail}
\end{figure}

\paragraph{Obs. 10. Available upstream results do not always lead to downstream results.}
Figure~\ref{fig:dependency-detail} shows which side of each dependency is missing. AOrchestra produces all 48 housing allocations but no handover records. Meta-Team produces market eligibility decisions and pitch allocations, while the corresponding handovers and inspection arrangements are missing. Only upstream records are available in 22.2\% and 16.7\% of their dependency outcomes, respectively, averaged by task. Superagent records venue allocations for all 42 exhibition objects without a transport record. Its volunteer task also has 20 pre-existing training bookings without matching calendar entries. Across its dependency checks, 22.2\% of outcomes have only the upstream record and 77.8\% have neither required output.

\paragraph{Obs. 11. Result coverage and cross-agent consistency reveal different failures.}
The market task illustrates the distinction. AOrchestra produces comparable allocation and inspection records for all 48 inspections, but 39 disagree; Meta-Team leaves the inspection side unavailable. In student support, AOrchestra has seven consistent eligibility--handover pairs out of 96, whereas Meta-Team has all 96. Counting consistency only where both results are available would hide this coverage difference. Contributor tracing further identifies six fully comparable relations handled across agents: three for AOrchestra and three for Meta-Team. Five contain inconsistencies, while Meta-Team's conservation schedule and staffing agree for all 32 group placements. Thus, completing related outputs and coordinating them are distinct requirements exposed by the task structure.

\begin{table}[t]
\centering\small
\setlength{\tabcolsep}{3pt}
\renewcommand{\arraystretch}{1.13}
\begin{tabularx}{\columnwidth}{@{}>{\raggedright\arraybackslash}Xrrr@{}}
\toprule
\textbf{Cross-agent relation} & \textbf{Items} & \multicolumn{2}{c}{\textbf{Inconsistent}} \\
\cmidrule(lr){3-4}
 & & \textbf{First} & \textbf{Final} \\
\midrule
\multicolumn{4}{@{}l}{\textit{AOrchestra}} \\
Radio rights $\to$ text & 24 & 24 & 24 \\
Market eligibility $\to$ handover & 84 & 72 & 72 \\
Market allocation $\to$ inspections & 48 & 39 & 39 \\
\midrule
\multicolumn{4}{@{}l}{\textit{Meta-Team}} \\
Radio schedule $\to$ staffing & 12 & 12 & 12 \\
Conservation schedule $\to$ staffing & 32 & 0 & 0 \\
Training bookings $\to$ calendar & 20 & 11 & \textbf{6} \\
\bottomrule
\end{tabularx}
\caption{\textbf{Changes in cross-agent consistency.} First is the earliest state with comparable records for all listed items; Final is episode termination. The bold count reflects five corrected booking--calendar mismatches. Other rows are unchanged. Conservation counts 32 group placements across 16 sessions.}
\label{tab:graph-contributions}
\end{table}

\paragraph{Obs. 12. Correcting individual schedules can leave paired results inconsistent.}
We reconstruct changes to the six cross-agent relations from successful operations, starting from each episode's initial state. Table~\ref{tab:graph-contributions} compares the first state with comparable records for every item against termination. In volunteer training, the booking worker later moves five volunteers who can attend only the second week, reducing booking--calendar mismatches from eleven to six. All six remaining mismatches concern volunteers available in either week: each date is individually feasible, but the booking and calendar refer to different sessions. The other four inconsistent relations retain their mismatches at termination. These changes show why checking each branch's local requirements does not replace checking agreement across branches.

\FloatBarrier

\section{Additional Benchmark Validation (RQ3)}
\label{app:instantiation-diagnostics}

MAC-Bench requires both completed local work and consistency across dependent results and shared resources. Validation therefore examines whether these requirements are stated in the task materials, can be fulfilled in the environment, and are reflected in scoring. Specification review and reference evaluation cover all 48 tasks. Controlled changes on twelve tasks test whether scoring accepts valid alternatives and repairs while rejecting missing work, broken relations, and preservation violations. Further diagnostics trace these decisions to individual goals. Each goal is assessed by a group of checks; PS measures completed goals, while full resolution also requires applicable preservation checks.

\subsection{Task Specifications}
\label{app:local-evidence}
\paragraph{Requirements and scoring checks.}
We review the 283 goals across all 48 test tasks. For each goal, we identify its public requirement, the records or artifacts needed to fulfill it, and the checks used to award credit. Business tasks provide source records, policies, and recording guides; software tasks provide interface specifications, starter code, and public tests. Preservation requirements are reviewed alongside the requested changes. For software, the review examines selected behavior groups and targeted assertions against the supplied interfaces and specifications.

We check this correspondence in both directions. Starting from a requirement, we examine whether the grader checks the requested behavior and its dependencies. Starting from a check, we locate the public basis for its rejection conditions, including permitted alternatives, required formats, and protected records or behavior. The resulting review records link each outcome to public source passages and grading entrypoints.

\paragraph{Review examples.}
The review checks both conditional requirements and the availability of required outputs. In the riverbank activity-planning task, notices must match the chosen session and its site's restrictions. We corrected a check that required a nesting-zone exclusion at two sites where none applied; it now tests the selected site's restrictions. In the university support-allocation task, a required and graded budget summary was missing from the structured list of outputs. We added its entry to the recording interface.

\subsection{Reference Solutions}
\label{app:native-references}
\paragraph{Execution procedure.}
Business reference solutions begin from the task's initial application state and use public tools to create or update the required records. We export the resulting state, apply the complete task grader, and reset the environment to check that the initial records are restored. Software reference solutions are installed in the supplied task environments, built with their prepared dependencies, and evaluated by the complete task grader, including checks of retained behavior. We separately evaluate the untouched application states and public software workspaces.

Reference results pass on 48/48 tasks, with eight per category; no untouched initial state passes complete grading. All 24 software starters build successfully before their initial-state evaluation. All business environments pass reset checks after reference execution and are closed after evaluation.

\paragraph{Materials and execution records.}
Evaluation records link each task and grader version to the evaluated state or source artifact and its check and goal scores. Business records include the public-tool trace and exported final state. These supplementary checks use the revised release packages; the main-system evaluation uses the packages from its original runs. Housing and screening grader corrections are evaluated on the unchanged final states from their reference executions.

\subsection{Scoring under Controlled Changes}
\label{app:software-relationship-checks}
\paragraph{Task selection and candidates.}
We select two tasks per category, covering distinct requirements such as financial reconciliation, allocation and handover, publication, and interacting software components. For each task, we construct six candidate results from its reference final state or implementation (Table~\ref{tab:validation-conditions}) and submit them to the grader, yielding 72 evaluations. Each candidate specifies the expected task decision; business candidates also specify which goals should pass or fail.

\begin{table}[t]
\centering\small
\renewcommand{\arraystretch}{1.12}
\setlength{\tabcolsep}{3pt}
\begin{tabularx}{\columnwidth}{@{}l>{\raggedright\arraybackslash}X@{}}
\toprule
\textbf{Condition} & \textbf{Change to the evaluated result} \\
\midrule
\rowcolor{black!5}\multicolumn{2}{@{}l}{\strut\textit{Expected to pass complete grading}} \\
Reference & Complete result from the reference solution. \\
Legal alternative & A different permitted choice or implementation, with dependent results updated. \\
Repair & Correct the inconsistent business records or restore the edited software component. \\
\midrule
\rowcolor{black!5}\multicolumn{2}{@{}l}{\strut\textit{Expected to fail complete grading}} \\
Missing work & Remove a required result register or omit required functionality. \\
Broken relation & Leave a consumer inconsistent with an upstream result or configuration. \\
Preservation violation & Alter protected source evidence or retained software behavior. \\
\bottomrule
\end{tabularx}
\caption{\textbf{Controlled scoring protocol.} Each of twelve tasks is evaluated under all six conditions. The expected decision follows the task requirements; Table~\ref{tab:verifier-calibration} reports observed scores for the violations.}
\label{tab:validation-conditions}
\end{table}

\paragraph{Candidate construction.}
Legal alternatives change a permitted choice or implementation while retaining the required behavior. Business alternatives select another allocation, schedule, menu, or session and update dependent records. For example, a housing reallocation includes a matching handover. In Lark, successive stable sorting passes replace a tuple-key sort while preserving the required ordering.

Relation violations leave connected results inconsistent: housing handover retains the old unit, Build omits the requested backend configuration, and label or role changes in FreshRSS and Kanboard leave old associations in place. Preservation violations alter protected source notes or existing software behavior, such as Jinja's exception line number. Business repairs synchronize dependent records with the permitted upstream choice; software repairs restore the edited component to its reference implementation.

\paragraph{Evaluation procedure.}
Business candidates are evaluated by the task's SQL checks and goal aggregation. Software candidates undergo a fresh build and complete task grading. We record task resolution, goal decisions, preservation checks, and PS. Table~\ref{tab:verifier-calibration} reports the scores under the three types of violation.

\paragraph{Obs. 13. Valid alternatives pass, while inconsistent results fail.}
All twelve references, twelve legal alternatives, and twelve repaired results pass with PS of 100\%. All 36 violations fail complete grading. In the business cases, the goal-level results also match the specified affected and retained outcomes across all 36 candidates. The following diagnostics examine how the scores respond when errors affect a dependent result, a shared resource, or preserved behavior.

\subsection{Relation and Goal-level Diagnostics}
\label{app:task-relationships}
Changing one branch's output can invalidate downstream work; individually valid choices can also conflict over a shared resource. We test whether goal-level scores capture these consequences while retaining credit for unaffected work.

\paragraph{Dependency consistency and completeness.}
The housing task assigns households to units and requires allocation, finance, appointment, and handover records. A permitted change in allocation must be reflected in the household's handover. Retaining the old unit in the handover fails that goal while the other three remain complete, giving PS of 75\%; synchronizing the handover restores full acceptance.

We also remove each upstream register separately from a complete result. This tests whether downstream checks require their supporting records to exist. The diagnostic exposed a missing-record case in the handover check, which we corrected with explicit per-household presence checks. Table~\ref{tab:housing-dependency-diagnostic} reports the resulting goal failures. Removing allocation also invalidates appointments, whereas removing finance or appointments leaves the other upstream goals complete. All three removals fail the handover goal; the reference continues to pass.

\begin{table}[t]
\centering\small
\renewcommand{\arraystretch}{1.15}
\setlength{\tabcolsep}{2pt}
\begin{tabular*}{\columnwidth}{@{\extracolsep{\fill}}lccccr@{}}
\toprule
\multirow{2}{*}{\textbf{Condition}} & \multicolumn{4}{c}{\textbf{Goal passed}} & \textbf{PS} \\
\cmidrule(lr){2-5}
& Alloc. & Fin. & Appt. & Hand. & (\%) \\
\midrule
Reference & \checkmark & \checkmark & \checkmark & \checkmark & 100 \\
Stale handover & \checkmark & \checkmark & \checkmark & $\times$ & 75 \\
Repaired handover & \checkmark & \checkmark & \checkmark & \checkmark & 100 \\
\midrule
\rowcolor{black!5}\multicolumn{6}{@{}l}{\strut\textit{Remove an upstream register from the reference}} \\
Allocation & $\times$ & \checkmark & $\times$ & $\times$ & 25 \\
Finance & \checkmark & $\times$ & \checkmark & $\times$ & 50 \\
Appointments & \checkmark & \checkmark & $\times$ & $\times$ & 50 \\
\bottomrule
\end{tabular*}
\caption{\textbf{Goal-level diagnosis in housing allocation.} Columns show allocation, finance, appointments, and handover. A stale handover fails its own goal; missing upstream records also invalidate dependent goals. $\checkmark$ and $\times$ indicate pass and fail.}
\label{tab:housing-dependency-diagnostic}
\end{table}

\paragraph{Shared-resource conflicts.}
We test shared-resource constraints in university support allocation and riverbank activity planning. Each has four candidate results: the reference, a conflicting assignment, a feasible swap, and restoration of the original arrangement. University support adds a thirteenth task to the calibration coverage. Two students can each use either of two remote-interpretation services, each with one place available for the term. Assigning both to one service exceeds its capacity. In riverbank activities, two overlapping sessions are assigned the same equipment kit. A feasible swap removes each conflict while maintaining individual suitability.

\paragraph{Obs. 14. Partial scores follow the affected requirements.}
The university conflict fails the allocation goal while preserving credit for five other goals, changing PS from 100\% to 83.3\%. The riverbank conflict fails the resources goal while four others remain complete, yielding PS of 80\%. Both swaps restore PS to 100\% and complete acceptance. The task-level and goal-level decisions match the specified outcomes in all eight conditions.

Software checks show how different errors affect different sets of goals. Ignoring Lark's token-retention option fails the tree goal while the other four goals pass, giving PS of 80\%. In Jinja, leaving bound macro arguments marked as unused fails the language and composition goals, while contexts, asynchronous execution, and loading retain credit, giving PS of 60\%. Altering Jinja's exception line number instead preserves all five task goals but fails the retained-behavior check: PS is 100\%, yet the task is rejected. Thus the score records completed goals, while full acceptance additionally requires preservation constraints to hold.

\paragraph{Application-state transitions.}
In the health-screening task, cancellation requires refunding affected payments. The reference execution exposed a payment-lifecycle mismatch: after a completed refund, Pretix marks the original payment as \texttt{refunded}, while the check required \texttt{confirmed}. We corrected the check to recognize the completed-refund state while requiring exactly one completed refund for the original payment and amount. Regrading the unchanged recorded final state passes all 95 individual verifier checks. Six controlled variants---missing, pending, duplicate, wrong-amount, incorrectly linked, and unpaid-source refunds---all fail the refund requirement and complete grading. This checks the legitimate state transition together with its corresponding error cases.

\FloatBarrier

\setcounter{topnumber}{2}
\setcounter{totalnumber}{3}
\section{Task Construction Analysis (RQ3)}
\label{app:additional-experiments}

MAC-Bench constructs tasks by reusing and recombining structures of interdependent work. We examine whether local patterns and connecting rules recur across tasks, whether composition expands the range of structures meeting specified requirements, and how target guidance and acceptance criteria affect these results. Finally, we compare the scenario specifications produced by alternative instantiation procedures.

\subsection{Cross-task Structure Reuse}
\label{app:structure-reuse}
\begin{figure}[t]
\centering
\includegraphics[width=\columnwidth]{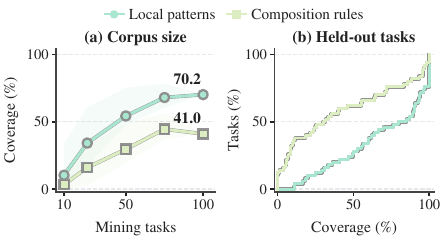}
\caption{\textbf{Task-structure reuse on 50 held-out tasks.} (a) Coverage versus corpus size, with ranges across ten nested sequences. (b) Coverage distributions with 100 mining tasks.}
\label{fig:pattern-coverage}
\end{figure}
Local patterns describe dependencies and coupling among two or three typed subtasks; connecting rules describe how two patterns overlap through shared subtasks. To examine their reuse beyond the mining corpus, we extract libraries from up to 100 tasks and measure coverage on 50 held-out tasks from disjoint source groups. A local occurrence is covered when its subtask types and relations match a library pattern. A connection is covered when both pattern types and their shared-node alignment match a stored rule. We compute each coverage fraction within a task, then average across tasks. Ten nested mining sequences use five corpus sizes, yielding 41 distinct libraries with a shared 100-task endpoint (Appendix~\ref{app:pattern-coverage-details}).

\paragraph{Obs. 15. Larger mining corpora cover more held-out structures.}
As shown in Figure~\ref{fig:pattern-coverage}(a), increasing the mining corpus from 10 to 100 tasks raises local-pattern coverage from 10.3\% to 70.2\% and rule coverage from 3.4\% to 41.0\%. Source-group bootstrap intervals for the gains are [53.3,66.2] and [27.4,49.1] percentage points. Rule coverage is 44.6\% at 75 tasks; subtask types are re-extracted at each corpus size, so coverage can decrease between sizes.

\paragraph{Obs. 16. Local patterns transfer more broadly than their composition rules.}
Figure~\ref{fig:pattern-coverage}(b) complements the means with task-level distributions. Median coverage is 82.5\% for local patterns and 30.8\% for rules; interquartile ranges are 49.8--96.9\% and 7.3--72.1\%. Twelve tasks have full local-pattern coverage, but only three have full rule coverage, and five have no rule match. Thus, recurring local configurations are more widely represented than the particular ways they connect into larger structures.

\paragraph{Coverage by source and relation type.}
\label{app:coverage-breakdown}
Table~\ref{tab:source-coverage} reports mean task-level coverage by source under the full library. The LHTB entry represents its single held-out task.

\begin{table}[t]
\centering\small
\setlength{\tabcolsep}{4pt}
\renewcommand{\arraystretch}{1.12}
\begin{tabularx}{\columnwidth}{@{}>{\raggedright\arraybackslash}Xrrr@{}}
\toprule
{\bfseries Held-out source} & {\bfseries Tasks} & {\bfseries Local} & {\bfseries Rules} \\
\midrule
MBABench & 9 & 57.7 & 22.7 \\
UniPat SaaS-Bench & 5 & 23.8 & 2.2 \\
RoadmapBench & 15 & 72.9 & 34.9 \\
SaaSBench & 9 & 96.7 & 79.4 \\
SWE-Milestone & 11 & 76.4 & 51.2 \\
LHTB & 1 & 70.0 & 33.3 \\
\midrule
\rowcolor{gray!10}All tasks & 50 & 70.2 & 41.0 \\
\bottomrule
\end{tabularx}
\caption{Held-out structure coverage by source (\%). Each value averages task-level coverage; All tasks averages the 50 tasks rather than the source means.}
\label{tab:source-coverage}
\end{table}

A post-hoc breakdown attributes 70.2\% of local-pattern occurrences to covered types, 8.3\% to types observed below the support threshold, 21.4\% to observed topologies with unrepresented type combinations, and 0.1\% to unseen topologies (task-macro shares). Coupling-containing patterns attain 67.2\% coverage on 25 eligible tasks; patterns without coupling attain 76.6\% on 49. Each fraction is computed over occurrences of the corresponding pattern class within eligible tasks.

\subsection{Controlled Structure Composition}
\label{app:structural-expansion}
\begin{figure}[t]
\centering
\includegraphics[width=\columnwidth]{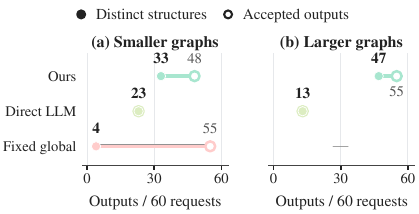}
\caption{\textbf{Accepted outputs and distinct structures.} Filled and open markers show distinct accepted structures and all accepted outputs from 60 requests, respectively. Their gap counts repeated structures; the remainder fails checks. No eligible larger fixed-global source exists (---).}
\label{fig:construction-yield}
\end{figure}

Coarse-to-fine composition changes the connections among subtask groups, then expands each group with local patterns. We test whether this produces distinct task graphs beyond the existing global structures. The comparison includes composition, Direct LLM generation, and a fixed-global reference that expands local patterns while keeping an existing global structure. All use the same library extracted from 100 tasks, targets, and checks. Each applicable condition receives 60 requests at each of two graph scales; Direct LLM uses one call per request without repair.

The targets specify how many terminal subtask groups the coarse graph contains and how many integrate outputs from multiple upstream groups. Acceptance also requires sufficient leaf-level work connected through explicit artifacts or shared-resource accesses (Appendix~\ref{app:controlled-generation-details}). Figure~\ref{fig:construction-yield} separates accepted outputs from distinct accepted structures. Composition produces fifteen repeated accepted structures at the smaller scale and eight at the larger scale; every accepted Direct LLM output is distinct.

For structural comparison, we remove names and IDs while retaining group categories, typed interfaces, directed dependencies, coupling, and shared-resource participation. Labeled graph isomorphism determines whether two outputs have the same structure. Weisfeiler--Lehman signatures and graph invariants accelerate comparisons by separating outputs that cannot be isomorphic.

\paragraph{Obs. 17. Composition expands distinct organizations meeting the chosen requirements.}
The fixed-global reference expands local patterns within four eligible source structures without global operations. It accepts 55 smaller-scale requests, yielding four organizations; none meets the larger target. Composition yields 33 distinct accepted organizations at the smaller scale and 47 at the larger scale, including 29 smaller-scale structures absent from comparable source structures.

\begin{figure}[t]
\centering
\includegraphics[width=\columnwidth]{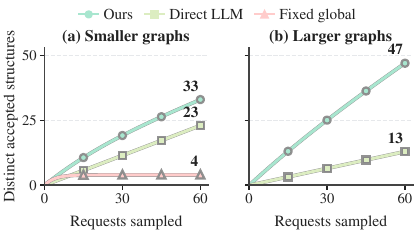}
\caption{Expected discovery of distinct accepted structures within each recorded 60-request batch. Each curve is computed from uniformly sampled subsets of the recorded requests, including failed requests.}
\label{fig:compositional-expansion}
\end{figure}

Figure~\ref{fig:compositional-expansion} shows the expected number of distinct accepted structures in a uniformly sampled subset of each 60-request batch. Among accepted composition outputs, 33 of 48 are distinct at the smaller scale and 47 of 55 at the larger scale; all accepted Direct LLM outputs are distinct. The curves combine acceptance frequency with repetition among accepted outputs.

\subsubsection{Target Guidance and Operation Limits}
\label{app:composition-diagnostics}
Target guidance selects refinement, integration, and coupling operations that move the coarse graph toward the requested group and integration counts, stopping when the targets are met or no improving update remains. We remove this guidance while retaining the same operation cap and local expansion routine. Acceptance under the resource-backed criterion drops from 80.0\% to 5.0\% at the smaller scale and from 91.7\% to 26.7\% at the larger scale, with no reduction in recorded request time (Table~\ref{tab:target-guidance}).

\paragraph{Operation-limit sensitivity.}
\label{app:operation-budget}

\begin{figure}[t]
\centering
\includegraphics[width=\columnwidth]{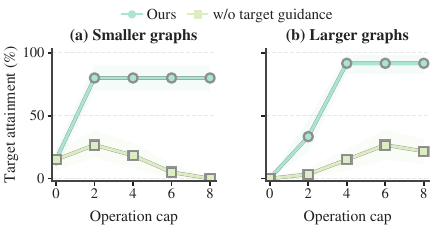}
\caption{\textbf{Effect of the global operation cap on target attainment.} Each point uses 60 paired requests; bands show 95\% within-seed request-bootstrap intervals. The resource-backed check and target scales are the same as in Figure~\ref{fig:construction-yield}.}
\label{fig:operation-budget}
\end{figure}
We use the same 100-task structure library, both target scales, and 60 requests per scale, setting global operation caps to 0, 2, 4, 6, and 8. Paired requests share initialization and local random streams, while local expansion and resource checks remain fixed. The unguided variant removes both target-directed selection and target/positive-gain stopping. The zero-operation condition still performs source initialization and local expansion; Fixed global instead preselects eligible sources.

The paired operation-cap protocol and uncertainty estimates are specified in Appendix~\ref{app:controlled-generation-details}.

\paragraph{Obs. 18. Target guidance matters beyond the operation count.}
As shown in Figure~\ref{fig:operation-budget}, guided composition reaches 80.0\% attainment after two operations at the smaller scale and 91.7\% after four at the larger scale. Without guidance, smaller-scale attainment falls from 26.7\% at two operations to zero at eight; larger-scale attainment peaks at 26.7\% at six. More operations alone do not ensure target attainment. Guided plateaus partly reflect stopping once targets are met.

\subsubsection{Acceptance Criteria and Generation Failures}
\label{app:acceptance-sensitivity}

\begin{figure}[t]
\centering
\includegraphics[width=\columnwidth]{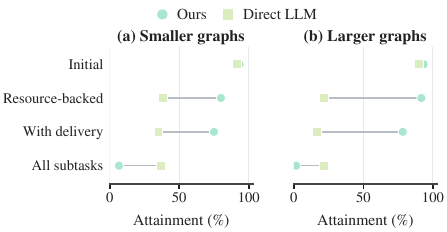}
\caption{Sensitivity of target attainment to the acceptance criterion. All four checks use the same 60 requests per method and target scale. Resource-backed, With delivery, and All subtasks denote the criteria in Appendix~\ref{app:resource-audit}.}
\label{fig:resource-audit-sensitivity}
\end{figure}

The task design calls for substantial branches connected through intermediate outputs or shared resources. The resource-backed criterion assesses these properties in generated graphs: work counts toward the required minimum when it has explicit artifact or shared-resource links, and integration requires artifacts from distinct upstream groups. We compare this criterion with the initial structural check and two variants on the same requests (Figure~\ref{fig:resource-audit-sensitivity}); Appendix~\ref{app:resource-audit} gives their thresholds.

\paragraph{Obs. 19. Generator comparisons depend on the structural criterion.}
The initial criterion counts groups and their declared connections. The two variants extend the resource-backed check by including final-delivery groups in the integration count or requiring explicit resource support for every subtask. The latter strengthens a minimum amount of interconnected work into a requirement on all subtasks.

Requiring every subtask to have resource support reverses the generator ranking: composition accepts 4 of 60 smaller-scale outputs and 1 of 60 larger-scale outputs, compared with 22 and 13 for Direct LLM. Including delivery groups can also change acceptance because it increases the integration count while retaining its upper bound.

Under the initial criterion, composition accepts 56 of 60 outputs at both scales; Direct LLM accepts 55 and 54 at the smaller and larger scales, respectively. Direct LLM's failed submissions include six transport failures and one truncated response. All submissions remain in the denominator. Diagnostic labels cover transport, schema, interface, target, and resource-support failures; one output may receive multiple labels.

\subsubsection{Generation Resources}
\label{app:generation-cost}
\paragraph{Obs. 20. Library-based composition avoids further model calls at the graph stage.}
After structure extraction, composition uses no further model calls and takes a mean 0.071 and 0.212 seconds per graph request at the smaller and larger scales. Direct LLM takes 63.69 and 75.93 seconds, respectively (Table~\ref{tab:generation-resources}). These times cover graph generation under the recorded execution configuration, excluding mining, development, and task instantiation. Token usage is available for 114 of 120 Direct LLM submissions; the six transport failures have no returned usage record.

\begin{table}[t]
\centering\small
\setlength{\tabcolsep}{3pt}
\renewcommand{\arraystretch}{1.12}
\begin{tabularx}{\columnwidth}{@{}>{\raggedright\arraybackslash}Xrrrr@{}}
\toprule
{\bfseries \multirow{2}{*}{Generator}} & \multicolumn{2}{c}{\bfseries Smaller graphs} & \multicolumn{2}{c}{\bfseries Larger graphs} \\
\cmidrule(lr){2-3}\cmidrule(l){4-5}
 & {\bfseries Pass (\%)} & {\bfseries Time (s)} & {\bfseries Pass (\%)} & {\bfseries Time (s)} \\
\midrule
Composition & 80.0 & 0.071 & 91.7 & 0.212 \\
Direct LLM & 38.3 & 63.69 & 21.7 & 75.93 \\
w/o guidance & 5.0 & 0.347 & 26.7 & 0.352 \\
\bottomrule
\end{tabularx}
\caption{Target attainment and recorded mean graph-generation time, with 60 requests per generator and scale. Pass uses the resource-backed criterion; time excludes mining and instantiation.}
\label{tab:generation-resources}
\label{tab:target-guidance}
\end{table}

\subsection{Exploratory Scenario Comparison}
\label{app:scenario-comparison}
During environment-grounded instantiation, abstract subtasks become concrete work with entities, outputs, and shared resources supported by the available applications. We compare the resulting scenario specifications under three procedures: refining composed graphs, refining existing source structures, and generating scenarios directly. Thirty business backgrounds each receive two requests per method, yielding 180 attempts. The methods share backgrounds, application capabilities, requirements for branches and their relations, output schema, and single-call budgets.

Programmatic checks assess schema, identifier consistency, dependency acyclicity, artifact bindings, and size. DeepSeek-Flash at temperature zero then evaluates each specification against common collaboration and environment requirements, with method names hidden. Each judgment must cite exact passages from the generated specification. Acceptance requires all criteria to be supported by valid citations; failed requests and unsupported judgments count as failures.

\begin{table}[t]
\centering\small
\renewcommand{\arraystretch}{1.12}
\begin{tabularx}{\columnwidth}{@{}>{\raggedright\arraybackslash}Xr@{}}
\toprule
{\bfseries Construction} & {\bfseries Acceptance (\%)} \\
\midrule
Ours & 40.0 \\
Source + refinement & 40.0 \\
Direct LLM & 46.7 \\
\bottomrule
\end{tabularx}
\caption{Acceptance of generated scenario specifications under common structural checks and automatic review supported by specification excerpts. Each condition uses 60 requests across 30 paired backgrounds.}
\label{tab:scenario-refinement}
\end{table}

\paragraph{Obs. 21. Observed specification acceptance rates are close.}
Accepted specifications number 24/60, 24/60, and 28/60 (Table~\ref{tab:scenario-refinement}). Paired-background bootstrap differences are 0.0 points versus source refinement (95\% interval $[-16.7,16.7]$) and $-6.7$ versus direct generation ($[-21.7,8.3]$). Both intervals include zero, leaving the direction of the acceptance differences unresolved.

\FloatBarrier

\setcounter{topnumber}{4}
\setcounter{totalnumber}{6}
\section{Task Construction Details}
\label{app:method-details}

We detail the three construction stages in Section~\ref{sec:benchmark-construction}. Mining turns existing tasks into reusable local patterns and global structures; composition combines them into new task graphs; instantiation specifies executable tasks in supported environments. Corpus statistics and comparison-specific candidate limits appear in Appendices~\ref{app:benchmark-statistics} and~\ref{app:experimental-details}; validation results appear in Appendix~\ref{app:instantiation-diagnostics}.

\subsection{Dependency-aware Task Structure Mining}
\label{app:structure-learning}

\paragraph{Task graph abstraction.}
A task specification and its accompanying documents, data, or code supply the evidence for recursive decomposition. Starting from $\mathcal T_i$, the policy $\pi_{\mathrm{dec}}$ proposes child subtasks, each described by its goal, inputs, outputs, and shared-resource accesses $\langle g_t,i_t,o_t,a_t\rangle$. Review checks whether their outcomes cover the parent's objective, whether required external inputs and outputs are retained, and whether the children specify distinct local results. Recursion ends at a coherent local result or when no acceptable further decomposition is obtained within the attempt limit. The retained tree is $\mathcal H_i$, with leaf set $\mathcal U_i$.

We infer relations between leaf subtasks from their inputs, outputs, and resource accesses. An output used as another leaf's required input, or a write establishing its required resource condition, supports a directed dependency. Coupling identifies mutually consequential modifications to a shared resource; sharing a read-only document or using unrelated records in one application is insufficient. The graph $\mathcal G_i$ carries directed edges $E_i^{\mathrm{dep}}$ and separate coupling annotations $E_i^{\mathrm{cpl}}$, following Equation~\ref{eq:work-graph}. Relation records retain artifact or resource identity, access modes, and supporting task evidence. Checks verify that referenced subtasks and resources exist, linked interfaces are compatible, and directed dependencies are acyclic. The recorded evidence supports the inferred meaning of each relation.

\paragraph{Local collaboration patterns: subtask types.}
Cross-task matching first assigns leaf subtasks to types $\tau$. Four distances compare objective semantics, interfaces, hierarchy context, and graph position. Semantic distance uses cosine dissimilarity of objective embeddings. Interface distance compares sets of input/output and resource-access descriptors using Jaccard dissimilarity. Hierarchy distance uses shared ancestry within a task; leaves from different task trees have no shared ancestry. Graph distance combines differences in structural position (such as source, sink, or fan-in), input/output behavior, and resource-access mode with normalized in/out-degree differences.

Let $\mathcal U_{\mathrm{all}}$ pool leaves from the mining tasks, with task indices distinguishing node identities. Let $b$ index these four distances $\delta_b(t,t')$, with fixed weights $\omega_b$, and let $k$ be the candidate number of subtask types. We choose a set of representative leaves, or medoids, $\mathcal J$ by minimizing weighted distance:
\begin{equation}
\begin{aligned}
\Delta(t,t')&=\sum_b\omega_b\delta_b(t,t'),\\
\mathcal J^\star&=\arg\min_{\substack{\mathcal J\subseteq\mathcal U_{\mathrm{all}}\\|\mathcal J|=k}}
\sum_i\frac{1}{|\mathcal U_i|}\sum_{t\in\mathcal U_i}
\min_{c\in\mathcal J}\Delta(t,c).
\end{aligned}
\label{eq:appendix-role-learning}
\end{equation}
For each candidate $k$, we alternate nearest-medoid assignment and weighted medoid updates, retaining the lowest-cost fit across initializations. Each medoid defines a type $\tau$, and every leaf receives its nearest medoid's type. The factor $1/|\mathcal U_i|$ gives each task equal total weight. For each type, we retain observed input/output descriptors and resource-access modes for later compatibility checks.

We select $k$ by maximizing cohesion minus weighted instability and low-support penalties. Cohesion is the fractional reduction in weighted distance relative to a single-medoid fit. Instability is the fraction of originally co-clustered leaf pairs separated after refitting with one task omitted, averaged over refits. The low-support penalty is the fraction of clusters represented in fewer than two tasks. Table~\ref{tab:mining-parameters} gives the candidate range, penalty weights, and refit settings.

\paragraph{Local collaboration patterns: canonicalization and connections.}
We enumerate connected two- and three-leaf sets, using dependency edges and coupling annotations for connectivity. Each occurrence retains its type labels and pairwise relations. The encoding distinguishes direct dependencies, indirect precedence established by reachability, incomparable nodes, and coupling. Thus, two contributing nodes in a fork or join can remain incomparable while belonging to one connected pattern. Equation~\ref{eq:pattern-canonicalization} gives the canonical encoding; identical encodings identify the same pattern type $M$ regardless of node names.

Within a source task, two pattern occurrences can share leaf subtasks. We record their pattern types $M$ and $M'$ and a mapping $\psi$ between the positions of the shared subtasks. The resulting rule $(M,M',\psi)$ identifies which nodes to merge when connecting these pattern types. For example, attaching a three-node pattern through one shared position introduces two new nodes. Each pattern and rule stores its supporting-task set; repeated occurrences within one task count once toward support. Both patterns and rules must meet their distinct-task support thresholds; a retained rule connects retained pattern types.

\paragraph{Global task structures and boundary mappings.}
For an internal node $h$ of the task tree $\mathcal H_i$, let $L_i(h)\subseteq\mathcal U_i$ be its descendant leaves. Each child of $h$ defines a group, and these groups partition $L_i(h)$. Contracting each child group gives $\overline{\mathcal G}_h$: a directed coarse edge exists when a leaf dependency crosses the corresponding groups, and coupling is retained when a coupling annotation crosses them. Internal relations stay inside their groups. The whole-task structure is extracted at the root; non-root internal groups provide refinement templates. Coarse cycles are excluded.

A boundary $\sigma(h)$ records externally supplied inputs, outputs needed outside the group or by the task objective, and shared-resource accesses involving other groups. Boundary mappings identify which child produces an output and which children consume an input or access a resource. For example, if a group must supply a report, the mapping identifies the child that produces that report after the group is expanded. A refinement template stores both the child-group graph and these mappings, allowing later replacement without losing external connections. Subtasks that combine multiple upstream contributions and feed further work supply integration templates; compatible shared-resource accesses supply coupling templates. The grouping tree records containment throughout, while coarse graphs record interactions.

\paragraph{Category indexing and library outputs.}
Group categories connect global structures to local pattern selection: a group's function determines which category-indexed patterns can expand it. The category $r$ describes that function: \texttt{foundation} supplies shared prerequisites, \texttt{capability} produces a local work result, \texttt{integrate} combines constituent work, and \texttt{deliver} provides the final task output. The experimental libraries assign these categories using deterministic rules over group objectives, decomposition, and boundary connections. For example, a root-level group that combines multiple upstream contributions and has no downstream group is categorized as \texttt{deliver}. These group categories index local patterns separately from the leaf-subtask types $\tau$.

A local occurrence is indexed under the smallest categorized tree group containing its leaves. We begin at their lowest common ancestor and move upward to the first categorized ancestor; occurrences with no categorized common ancestor are not included in the category-indexed library. For each category $r$, $\mathcal M_r$ stores its retained pattern types and $c_r(M)$ counts distinct supporting tasks. The local library $\mathcal L_L$ contains these indexed patterns, type interfaces, and overlap rules; the global library $\mathcal L_G$ contains coarse graphs, boundaries, and composition templates. These outputs form $\mathcal K=(\mathcal L_L,\mathcal L_G)$.

Algorithm~\ref{alg:structure-learning} summarizes this data flow. The retained abstractions form $\mathcal D$; $\Theta$ stores subtask types and their interfaces, and $\tau(t)$ is the type assigned to leaf $t$. $\mathcal O$ retains pattern occurrences with their task identities and node mappings, while $\mathcal A$ stores overlap rules. \textsc{Filter} applies the distinct-task support thresholds to patterns and rules. \textsc{Coarsen} extracts the global structures, boundaries, templates, and group categories described above; \textsc{Index} uses these categories to assemble $\mathcal L_L$.

\begin{algorithm}[tbp]
\caption{Extracting task structure libraries}
\label{alg:structure-learning}
\small
\begin{algorithmic}[1]
\Require Tasks $\{\mathcal T_i\}$ with accompanying materials
\Ensure $\mathcal K=(\mathcal L_L,\mathcal L_G)$
\State $\mathcal D\gets\varnothing$
\For{each task $\mathcal T_i$}
 \State $(\mathcal H_i,\mathcal U_i)\gets\Call{Decompose}{\mathcal T_i}$
 \State $\mathcal G_i\gets\Call{InferRelations}{\mathcal T_i,\mathcal U_i}$
 \If{$\Call{ValidAbstraction}{\mathcal H_i,\mathcal G_i}$}
  \State $\mathcal D\gets\mathcal D\cup\{(\mathcal H_i,\mathcal G_i,\mathcal U_i)\}$
 \EndIf
\EndFor
\State $(\Theta,\tau)\gets\Call{Cluster}{\mathcal D}$ \Comment{Eq.~\ref{eq:appendix-role-learning}}
\State $\mathcal O\gets\Call{Enumerate}{\mathcal D,\{2,3\}}$
\State $\mathcal O\gets\Call{Canonicalize}{\mathcal O,\tau}$ \Comment{Eq.~\ref{eq:pattern-canonicalization}}
\State $\mathcal A\gets\Call{Overlaps}{\mathcal O}$
\State $(\mathcal O,\mathcal A)\gets\Call{Filter}{\mathcal O,\mathcal A}$
\State $\mathcal L_G\gets\Call{Coarsen}{\mathcal D}$
\State $\mathcal L_L\gets\Call{Index}{\mathcal O,\mathcal A,\Theta,\mathcal L_G}$
\State \Return $(\mathcal L_L,\mathcal L_G)$
\end{algorithmic}
\end{algorithm}

\subsection{Coarse-to-fine Task Composition}
\label{app:task-composition}

\paragraph{Global structure composition.}
We initialize a whole-task coarse graph from $\mathcal L_G$ with its grouping tree, boundary interfaces, and categories. We assign fresh identifiers to groups and resources while retaining their types and internal correspondences. Refinement substitutes a group's stored child graph and reconnects its external interfaces through the boundary mapping. The replaced group becomes an internal tree node. Integration inserts a group consuming multiple upstream contributions and makes its output available to subsequent work. Coupling binds compatible accesses to a shared resource. Integration groups belong to the grouping tree, but dependency and coupling relations belong to the coarse graph, not to tree edges.

Let $\mathcal C=(\mathcal H,\overline{\mathcal G})$ be the current structure. Valid one-step candidates $\mathcal R(\mathcal C)$ preserve interfaces, resource-access compatibility, acyclicity, and the correspondence between terminal groups and coarse nodes. A template already applied in the current composition is not reused. For requested statistic intervals $[\ell_j,u_j]$, positive weights $w_j$, and statistics $z_j(\mathcal C)$, target deviation is
\begin{equation}
D(\mathcal C)=\sum_j w_j\max\{\ell_j-z_j(\mathcal C),0,z_j(\mathcal C)-u_j\}.
\label{eq:appendix-target-deviation}
\end{equation}
Each term is zero inside its requested interval and otherwise measures distance to the nearest bound. Group and integration counts are measured on the coarse graph; their experimental definitions appear in Appendix~\ref{app:controlled-generation-details}. At step $k$, $\mathcal C^{(k)}$ is the current structure. A nonnegative penalty $\rho_k(\mathcal C')$ favors unused template sources and discourages refinements with no internal bindings. We propose
\begin{equation}
\widehat{\mathcal C}\in\arg\min_{\mathcal C'\in\mathcal R(\mathcal C^{(k)})}
\bigl[D(\mathcal C')+\rho_k(\mathcal C')\bigr].
\label{eq:graph-composition}
\end{equation}
Ties are broken uniformly at random. We set $\mathcal C^{(k+1)}=\widehat{\mathcal C}$ only when $D(\mathcal C^{(k)})-D(\widehat{\mathcal C})>\rho_k(\widehat{\mathcal C})$. Search stops when the targets are met, no candidate has positive penalized gain, or the operation limit is reached. An output outside required intervals is rejected. The resulting $(\mathcal H^\star,\overline{\mathcal G}^\star)$ retains each terminal group's category, boundary, and local size target.

Algorithm~\ref{alg:task-composition} gives the composition procedure, with $B_G$ global updates, $B_R$ local attempts, and $B_L$ expansion steps per attempt. \textsc{Initialize} draws the starting structure from $\mathcal L_G$. \textsc{Apply} accepts a proposal and records its template and source for subsequent candidate filtering and penalties. \textsc{Expand} is detailed in Algorithm~\ref{alg:local-expansion}; pruning, boundary binding, and final checks follow the assembly procedure below. A failed construction returns $\bot$.

\begin{algorithm}[tbp]
\caption{Composing and assembling a task graph}
\label{alg:task-composition}
\small
\begin{algorithmic}[1]
\Require $\mathcal K$; target intervals; limits $B_G,B_R,B_L$
\Ensure $(\mathcal H^\star,\widetilde{\mathcal G})$ or $\bot$
\State $\mathcal C\gets\Call{Initialize}{\mathcal L_G}$
\For{$k=0,\ldots,B_G-1$}
 \State \textbf{if} $D(\mathcal C)=0$ \textbf{then break}
 \State $\mathcal Q\gets\mathcal R(\mathcal C)$
 \State \textbf{if} $\mathcal Q=\varnothing$ \textbf{then break}
 \State $\widehat{\mathcal C}\in\arg\min_{\mathcal C'\in\mathcal Q}[D(\mathcal C')+\rho_k(\mathcal C')]$
 \State \textbf{if} $D(\mathcal C)-D(\widehat{\mathcal C})\le\rho_k(\widehat{\mathcal C})$ \textbf{then break}
 \State $\mathcal C\gets\Call{Apply}{\mathcal C,\widehat{\mathcal C}}$
\EndFor
\State \textbf{if} $D(\mathcal C)>0$ \textbf{then return} $\bot$
\State $(\mathcal H^\star,\overline{\mathcal G}^\star)\gets\mathcal C$
\For{each terminal group $h$ in $\mathcal H^\star$}
 \State $\mathcal G_h\gets\Call{Expand}{h,\mathcal L_L,B_R,B_L}$
 \State $\mathcal G_h\gets\Call{Prune}{\mathcal G_h,\sigma(h)}$
\EndFor
\State $\widetilde{\mathcal G}\gets\Call{Bind}{\overline{\mathcal G}^\star,\{\mathcal G_h\},\mathcal L_L}$
\State $\widetilde{\mathcal G}\gets\Call{RemoveUnboundIsolates}{\widetilde{\mathcal G}}$
\State \textbf{if} $\Call{Check}{\widetilde{\mathcal G}}$ \textbf{then return} $(\mathcal H^\star,\widetilde{\mathcal G})$
\State \Return $\bot$
\end{algorithmic}
\end{algorithm}

\paragraph{Local pattern composition.}
For a terminal group $h$ of category $r$, we draw a size-compatible seed from $\mathcal M_r$ using Equation~\ref{eq:motif-sampling}, renormalized over eligible candidates. Interface coverage is checked over the growing group, so several patterns can jointly supply its required inputs, outputs, and resource accesses. A group with a single-subtask target uses a subtask type observed in that category, preferring one that covers its boundary.

For larger groups, we retain each placed pattern instance and its mapping from pattern positions to graph nodes. At each step, we uniformly select an instance with at least one eligible extension. Let $M$ be its pattern type, $\Psi(M)$ the stored extensions $(M',\psi)$, and $c_M(M',\psi)$ their distinct-task support. Conditional on this selected instance, the extension prior is
\begin{equation}
p(M',\psi\mid M)=\frac{c_M(M',\psi)+\beta}
{\sum_{(\bar M,\bar\psi)\in\Psi(M)}[c_M(\bar M,\bar\psi)+\beta]},
\label{eq:attachment-sampling}
\end{equation}
where $\beta>0$ smooths the counts. Eligible extensions must belong to category $r$'s retained pattern library, introduce new nodes, preserve shared-node types, stay within the size limit, and avoid directed cycles. We renormalize the prior over extensions eligible for the selected instance. Thus, an instance with more stored rules is not automatically more likely to be selected.

An attachment uses $\psi$ to map positions in the new pattern to existing nodes, then adds its remaining nodes, direct dependencies, and coupling annotations. The new instance joins the set eligible for subsequent extensions. Indirect precedence describes reachability in the source pattern; it is not copied as an additional direct edge. In the composed graph, precedence follows the assembled dependency paths.

Expansion stops when the local size target and interface coverage are reached, no eligible extension remains, or the step limit is reached. An unsuccessful attempt can restart from another seed within the attempt limit. The resulting $\mathcal G_h$ is then pruned and assembled; interface coverage and the requested graph-size requirements are checked on the resulting graph. Sampling uses distinct-task support without interface-coverage reweighting in the controlled experiments (Appendix~\ref{app:controlled-generation-details}).

In Algorithm~\ref{alg:local-expansion}, $n_h$ is the group's local size target, $V(\mathcal G_h)$ its node set, and $\mathcal I_h$ its placed pattern instances. Instance $I$ has pattern type $M_I$ and a position-to-node mapping. \textsc{Eligible} returns its available extensions $\mathcal E_h(I)$ under the conditions above; \textsc{Sample} renormalizes the prior over this set. \textsc{Attach} merges shared positions, inserts new nodes and relations, and records the new instance. \textsc{Covered} tests the growing graph against boundary $\sigma(h)$. Each attempt starts with \textsc{Seed}, which uses Equation~\ref{eq:motif-sampling}. Without an eligible seed, it returns the category's most frequent typed node with $\mathcal I_h=\varnothing$. For a single-subtask target, \textsc{Singleton} prefers a boundary-compatible type as described above.

\begin{algorithm}[tbp]
\caption{Local pattern expansion: \textsc{Expand}}
\label{alg:local-expansion}
\small
\begin{algorithmic}[1]
\Require Group $h$; local library $\mathcal L_L$; limits $B_R,B_L$
\Ensure Local graph $\mathcal G_h$
\State \textbf{if} $n_h=1$ \textbf{then return} $\Call{Singleton}{h,\mathcal L_L}$
\For{$a=1,\ldots,B_R$}
 \State $(\mathcal G_h,\mathcal I_h)\gets\Call{Seed}{h,\mathcal L_L}$
 \For{$s=1,\ldots,B_L$}
  \If{$|V(\mathcal G_h)|\ge n_h$ \textbf{and} $\Call{Covered}{\mathcal G_h,\sigma(h)}$}
   \State \textbf{break}
  \EndIf
  \State $\mathcal E_h\gets\Call{Eligible}{\mathcal G_h,\mathcal I_h,h,\mathcal L_L}$
  \State $\mathcal I_h^+\gets\{I\in\mathcal I_h:\mathcal E_h(I)\ne\varnothing\}$
  \State \textbf{if} $\mathcal I_h^+=\varnothing$ \textbf{then break}
  \State $I\sim\operatorname{Uniform}(\mathcal I_h^+)$
  \State $(M',\psi)\gets\Call{Sample}{p(\cdot\mid M_I),\mathcal E_h(I)}$
  \State $(\mathcal G_h,\mathcal I_h)\gets\Call{Attach}{\mathcal G_h,\mathcal I_h,I,M',\psi}$
 \EndFor
 \State \textbf{if} $\Call{Covered}{\mathcal G_h,\sigma(h)}$ \textbf{then break}
\EndFor
\State \Return $\mathcal G_h$
\end{algorithmic}
\end{algorithm}

\paragraph{Interface-constrained graph assembly.}
Before assigning interfaces, we prune each multi-node local graph around candidate boundary nodes and their dependency predecessors. Boundary nodes are selected for required inputs, outputs, or resource accesses; groups without an output boundary also retain local sinks. When these anchors can be identified, we retain them and their dependency predecessors. Single-node groups are retained as units.

Boundary assignment then turns coarse relations into concrete leaf-level bindings. For every required output, we select a compatible producer in $\mathcal G_h$. Required inputs are assigned to compatible consumers and linked to a producer or environmental supply. Shared-resource ports retain each participating node's read/write mode and require a writer where mutable state is involved. Compatibility is checked against the input/output types and resource-access modes stored in $\mathcal L_L$.

A cross-group dependency binds the upstream output to the downstream input, creating an explicit artifact relation between their assigned nodes. Shared-resource bindings align resource identity and access modes across groups, and local coupling annotations receive compatible resource accesses. These bindings can have multiple consumers or participants, so a group boundary need not map to a single node. Internal relations remain in the local graphs.

After binding, isolated nodes without an interface assignment are removed from multi-node groups. Size, boundary coverage, resource compatibility, and dependency acyclicity are then checked on the assembled graph $\widetilde{\mathcal G}$. Retaining a node through a dependency path does not require it to carry its own explicit artifact or resource binding. The resource-backed acceptance criterion counts only nodes with such bindings toward its work minimum; the stricter all-subtasks variant requires every node to qualify (Appendix~\ref{app:resource-audit}). The grouping tree $\mathcal H^\star$ is retained separately. Settings and comparison-specific candidate restrictions appear in Appendix~\ref{app:controlled-generation-details}.

\subsection{Environment-grounded Task Instantiation}
\label{app:task-instantiation}

\paragraph{Scenario refinement and task packaging.}
Given $(\mathcal H^\star,\widetilde{\mathcal G})$, refinement specifies concrete work in an environment $\Omega$ with suitable entities, artifacts, and permitted operations. Objectives and resource conditions are adapted to the available application records or project interfaces; work may be merged, split, or rearranged. Business scenarios record these requirements in mechanism specifications: descriptions of the entities, required work results, dependencies between results, and constraints on shared resources. Software tasks specify them through project requirements, public interfaces, and expected artifacts.

The output follows Equation~\ref{eq:task-grounding}: a description $d$, initial environment $x_0$, and verifier $V$. The task description states requested outcomes, constraints, and available evidence. Initial records or project files supply the starting resources and existing behavior. Private verification materials remain outside the solver workspace. The task package brings together these components and the permitted operations.

\paragraph{Outcome checks.}
Business checks inspect actual application records and designated result artifacts, including returned identifiers, dates, amounts, assignments, and applicable joint constraints. Software checks evaluate delivered implementations against required behavior and interfaces. Goal checks assess requested changes; preservation checks compare protected records or required existing behavior with the starting state. The verifier reports goal decisions and complete resolution, with the score definitions in Appendix~\ref{app:task-metrics}. It does not require a prescribed action sequence.

\paragraph{Validation of task packages.}
We review the correspondence between public requirements, available operations, and outcome checks, then test loading, reset, and grading in the packaged environment. Business reference solutions use public tools from the initial state and export their results for grading; software reference implementations are built and evaluated with the complete task grader. Constructed alternatives and violations test scoring decisions. Appendix~\ref{app:instantiation-diagnostics} details the review, reference results for all 48 tasks, and controlled scoring tests.

\section{Experimental Details}
\label{app:experimental-details}

\subsection{Baseline Implementation}
\label{app:baselines}
We adapt the five baselines to MAC-Bench's terminal and application-tool interfaces. Agents within each run share the environment and backbone; each system retains its own organization and coordination mechanism.

\paragraph{ReAct.}
ReAct \citep{yao2023react} uses one persistent decision--tool loop, with earlier interactions summarized to support long episodes.

\paragraph{AOrchestra.}
AOrchestra \citep{ruan2026aorchestra} retains its coordinator, delegated ReAct workers, and delegation history. Delegation is sequential; workers receive fresh contexts while accessing the shared task environment.

\paragraph{Meta-Team.}
Meta-Team \citep{hao2026metateam} uses planner, developer, and reviewer for software tasks, and a chairman with three workers for business tasks. Members maintain separate histories, and test-time reflection is disabled.

\paragraph{Magentic-One.}
Magentic-One \citep{fourney2024magenticone} retains task/progress ledgers, speaker selection, and stall-triggered replanning. Its software team comprises FileSurfer, Coder, and the non-model ComputerTerminal; its business team comprises Operator, Analyst, and Reviewer. Tool access is redirected to the task's permitted operations.

\paragraph{Superagent.}
Superagent \citep{deepseek2026superagent} uses a router to assign requests to workers through tickets and relay their reports. Workers plan and execute the work and can delegate to subagents. The router, workers, and subagents maintain separate conversation histories while sharing the task environment. We use the DeepSeek Harness for agent execution and context management, connecting agent tools to MAC-Bench's filesystem, shell, and application interfaces. Workers report completion, blockers, or failure through tickets; execution ends when the router and workers finish or an episode limit is reached.

\subsection{Evaluation Settings and Metrics}
\label{app:evaluation-settings}
\paragraph{Shared evaluation conditions.}
Each system--backbone--task combination is evaluated once from a reset environment with the same public instructions and permitted tools. Construction graphs, reference artifacts, and private verification materials are hidden. Development tasks support adaptation and budget selection before testing.

Task-specific token caps are assigned in advance and shared across systems and backbones: three tasks use 5M tokens, eleven use 10M, and thirty-four use 15M. Input and output tokens from all model calls, including subagents and summaries, count toward the same cap, including the full usage of an admitted final call. The wall-clock limit is two hours.

\paragraph{Model and context settings.}
GPT-6 Luna and DeepSeek-Flash each supply all generative roles within their corresponding runs, with thinking disabled. Tables~\ref{tab:baseline-decoding} and~\ref{tab:baseline-controls} list decoding and execution settings. ReAct and Magentic-One use rolling summaries, with a 131,072-token request limit, a 64,000-token input target, and a 12,288-token summary-output limit. They retain five recent turns and 20,000 recent tokens. AOrchestra summarizes at ten worker records and keeps the three most recent.

Superagent uses per-agent context compaction with configured windows of 1,050,000 tokens for GPT-6 Luna and 1,000,000 for DeepSeek-Flash. Automatic compaction triggers at 80\% of the window and summarizes older history; the recent-context retention target is 16\% of the input budget after reserving output tokens. Summary calls allow up to 65,536 output tokens. During compaction, tool results exceeding 8,192 characters retain their first 4,096 and last 1,024 characters.

\begin{table}[!t]
\centering
\small
\renewcommand{\arraystretch}{1.12}
\begin{tabularx}{\columnwidth}{@{}>{\raggedright\arraybackslash}Xrr@{}}
\toprule
{\bfseries Method} & {\bfseries Temperature} & {\bfseries Max. output tokens} \\
\midrule
ReAct & 0.0 & 8,192 \\
AOrchestra & 0.2 & 8,192 \\
Meta-Team & 0.2 & 32,768 \\
Magentic-One & 0.2 & 8,192 \\
Superagent & Default & 32,768 \\
\bottomrule
\end{tabularx}
\caption{Standard model-call decoding parameters for both backbones. Default denotes the provider's default temperature.}
\label{tab:baseline-decoding}
\end{table}

\begin{table}[!t]
\centering\small
\setlength{\tabcolsep}{4pt}
\renewcommand{\arraystretch}{1.1}
\begin{tabularx}{\columnwidth}{@{}>{\raggedright\arraybackslash}Xr@{}}
\toprule
\textbf{Control} & \textbf{Setting} \\
\midrule
\rowcolor{gray!10}\multicolumn{2}{@{}l}{\textbf{AOrchestra}} \\
Delegation & Sequential \\
Coordinator-attempt limit & 3,000 \\
Worker-step limit & 50 \\
Worker memory / recent records & 10 / 3 \\
\midrule
\rowcolor{gray!10}\multicolumn{2}{@{}l}{\textbf{Meta-Team}} \\
Member-step limit & 3,000 \\
Test-time reflection & Disabled \\
\midrule
\rowcolor{gray!10}\multicolumn{2}{@{}l}{\textbf{Magentic-One}} \\
Orchestrator-turn limit & 3,000 \\
Stall threshold & 3 \\
Worker tool iterations per turn & 10 \\
\midrule
\rowcolor{gray!10}\multicolumn{2}{@{}l}{\textbf{Superagent}} \\
Max. active subagents per worker & 8 \\
Max. subagent depth & 1 \\
Background-work wait limit & 30 min \\
\bottomrule
\end{tabularx}
\caption{Method-specific execution controls, shared by both backbones. Shared episode limits also apply.}
\label{tab:baseline-controls}
\end{table}

For ReAct, AOrchestra, Magentic-One, and Meta-Team, output-truncation recovery allows up to two additional calls. Regeneration can double the output limit up to 32,768 tokens for ReAct, AOrchestra, and Magentic-One, or 65,536 for Meta-Team. Recovery calls count toward the episode budget.

\paragraph{Scoring and resource accounting.}
\label{app:task-metrics}
Each required task goal is represented by a group of outcome checks. A goal is achieved only when all required checks in its group pass. Let $\mathcal Q_i$ be these goal groups for task $i$ and let $b_{iq}\in\{0,1\}$ indicate that group $q$ passes. For $n$ evaluated tasks, we report a task-normalized Progress Score and the fraction of fully resolved tasks:
\begin{equation}
\begin{aligned}
\mathrm{PS}_i &= \frac{1}{|\mathcal Q_i|}\sum_{q\in\mathcal Q_i}b_{iq},\\
\mathrm{RR} &= \frac{1}{n}\sum_{i=1}^{n}\operatorname{Resolved}(i).
\end{aligned}
\label{eq:appendix-task-scores}
\end{equation}
Here $\operatorname{Resolved}(i)\in\{0,1\}$ is the task's resolved flag. The evaluator's goal-completion fraction supplies $\mathrm{PS}_i$; the resolved flag additionally reflects the task's complete acceptance conditions, including preservation checks where applicable. We average PS across tasks and express both metrics as percentages. Business operations goal groups concern required native-state outcomes; Software Engineering groups concern the delivered project behavior and required checks. Fine-grained verifier checks support the goal decisions rather than being pooled across tasks as independent samples.

Construction and development costs are separate from task-solving costs.

\subsection{Analysis Protocols}
\label{app:analysis-protocols}
The task-level performance, resource-use, and collaboration analyses use the DeepSeek-Flash runs: 48 tasks per system for performance and resource use, and nine business tasks across five systems (45 episodes) for relation and contributor analysis.

\subsubsection{Preliminary Study}
\label{app:preliminary}
We uniformly sample 50 tasks from each of GAIA 2023 validation, SWE-bench Verified, and Terminal-Bench 2.0 using seed 42. Inputs include task instructions and available attachments or starter files, excluding reference answers, private tests, and execution traces. Repository and attachment excerpts have bounded text budgets.

A substantial branch pursues a separately checkable result and can make progress before exchanging intermediate outputs. Lookup steps and serial stages of a single repair are not separate branches. Interdependence requires an output dependency, consequential interface requirement, or shared-resource interaction; a common input alone is insufficient. The categories distinguish primarily serial work, independent branches, interdependent branches, shared-state coupling, and unresolved cases.

DeepSeek-Flash at temperature zero proposes structures. GPT-6 Astra High then reviews the task evidence and revises 52 of 150 labels. Both stages use the supplied task materials. Final counts for serial work, independent branches, interdependent branches, shared-state coupling, and unresolved cases are $(49,1,0,0,0)$ for GAIA, $(49,0,0,0,1)$ for SWE-bench Verified, and $(37,1,11,0,1)$ for Terminal-Bench. The two unresolved cases remain in the reported distribution.

\subsubsection{Collaboration Analysis}
\label{app:graph-collaboration-analysis}
We assess delivered results using fourteen dependency checks and six shared-resource checks (Table~\ref{tab:relation-diagnostic-scope}), then use execution traces to identify how agents divide and coordinate the related work.

\begin{table}[!t]
\centering
\small
\setlength{\tabcolsep}{3.2pt}
\renewcommand{\arraystretch}{1.12}
\begin{tabularx}{\columnwidth}{@{}>{\raggedright\arraybackslash}Xrrr@{}}
\toprule
{\bfseries Task} & {\bfseries Dependencies} & {\bfseries Shared} & {\bfseries Budget} \\
\midrule
Exhibition recovery & 1 & 0 & 15 \\
Anthology release & 2 & 0 & 15 \\
Radio release & 2 & 0 & 15 \\
Conservation programme & 2 & 1 & 15 \\
Membership renewal & 1 & 0 & 15 \\
Housing handover & 1 & 1 & 10 \\
Student support & 2 & 2 & 15 \\
Volunteer training & 1 & 0 & 15 \\
Market permits & 2 & 2 & 10 \\
\midrule
\rowcolor{gray!10}Total & 14 & 6 & \\
\bottomrule
\end{tabularx}
\caption{Tasks and relation checks in the collaboration analysis. Shared-resource checks include constraints among allocations within a work group. Budgets are in millions of tokens.}
\label{tab:relation-diagnostic-scope}
\end{table}

\paragraph{Relation assessment.}
For each dependency, we identify the upstream and downstream work and compare their corresponding outputs. In the volunteer-training task (Service Management), bookings and calendar entries must refer to the same training sessions. We match these records by volunteer identity and compare their dates. Matching dates are \textit{satisfied}; different dates are \textit{inconsistent}; missing required records are \textit{unavailable}; and records whose correspondence cannot be resolved are \textit{unassessable}. Shared-resource checks apply the same categories to joint constraints: complete arrangements are satisfied when they can coexist, and inconsistent when they conflict, such as overlapping use of the same equipment. Each check covers the people, activities, or allocations specified by the task, including those with missing results.

\paragraph{Relation outcomes.}
For each check, we compute the fractions of task-specified objects in the four outcome categories. We average these fractions over checks within each task, then across tasks, giving each task equal weight. Figure~\ref{fig:relation-outcomes} reports dependencies over nine tasks and shared-resource constraints over four. Across the five systems, the analysis comprises 4,380 object-level checks.

\paragraph{Agent contributions and coordination.}
Delegations, messages, and successful writes identify the contributors to each side of a task relation. Different single contributors indicate across-agent handling; the same contributor indicates same-agent handling. Multiple contributors indicate joint or changed authorship. For results stored together in a document, attribution follows that document's successful writers. Figure~\ref{fig:dependency-detail} retains every dependency outcome, including incomplete results and mixed attribution. Its stars identify the six relations with complete outcome evidence and distinct contributors on both sides for every item.

\paragraph{Availability and subsequent changes.}
For Figure~\ref{fig:dependency-detail}, we split unavailable outcomes into upstream-only, downstream-only, and neither-side availability. Availability means that the required records and fields can be matched for the relation check; it does not require the whole branch to pass its goal verifier. Initial records participate in the comparison alongside records created during execution. The fourteen checks cover 710 items per system, or 3,550 object-level comparisons in total.

For the six starred relations, we reconstruct relevant record changes from the initial snapshot and successful tool calls, retrieving full saved responses when event logs contain shortened responses. We apply changes in recorded completion order and compare the same objects at each step. The first comparable state is the earliest state at which every object has both sides available. Table~\ref{tab:graph-contributions} reports inconsistent objects at that point and at termination. Reconstructed final outcomes match the archived relation checks for every object.

\subsubsection{Held-out Structure Reuse}
\label{app:pattern-coverage-details}
We test whether extracted patterns and connecting rules match work structures in tasks excluded from mining. The 150-task corpus is divided into 100 mining tasks and 50 held-out tasks with 65 and 34 disjoint source groups. Tasks from the same repository or original case, and identical normalized specifications, remain together. Source counts appear in Table~\ref{tab:source-corpus}.

Ten source-stratified nested sequences use mining sizes 10, 25, 50, 75, and 100. Types, patterns, and connecting rules are extracted for each subset, yielding 41 distinct libraries and one shared full-corpus endpoint. Held-out leaves are assigned to the types extracted from that subset. Patterns and rules require support from at least two mining tasks.

A held-out pattern matches when its canonical encoding has the same subtask types and dependency and coupling relations as a library pattern. A connecting rule matches when the two pattern types and the mapping between their shared subtasks agree. For each measure, we divide matching occurrences by all corresponding occurrences within each held-out task, then average across the 50 tasks. Curve bands are minimum--maximum ranges across mining subsets. Bootstrap intervals use 2,000 resamples of the 34 held-out source groups, preserving within-group dependence.

\paragraph{Extraction settings.}
\label{app:mining-settings}
Extraction follows Appendix~\ref{app:method-details}; its parameters are collected in Table~\ref{tab:mining-parameters}.

\begin{table}[!t]
\centering\small
\setlength{\tabcolsep}{4pt}
\renewcommand{\arraystretch}{1.1}
\begin{tabularx}{\columnwidth}{@{}>{\raggedright\arraybackslash}Xr@{}}
\toprule
\textbf{Setting} & \textbf{Value} \\
\midrule
Objective embedding model & \texttt{BAAI/bge-m3} \\
Semantic / interface weights & 0.45 / 0.25 \\
Hierarchy / graph weights & 0.15 / 0.15 \\
Candidate subtask-type counts & 8--20 \\
Initializations / maximum iterations & 5 / 25 \\
Stability refits & 4 \\
Instability / low-support penalties & 0.35 / 0.40 \\
Minimum distinct-task support & 2 \\
\bottomrule
\end{tabularx}
\caption{Parameters for structure extraction. Distance weights follow Equation~\ref{eq:appendix-role-learning}.}
\label{tab:mining-parameters}
\end{table}

\subsubsection{Controlled Graph Composition}
\label{app:controlled-generation-details}
We compare generators under shared structural targets and acceptance checks. The comparison uses the 100-task libraries, excluding held-out tasks. Table~\ref{tab:learned-library} distinguishes extracted structures from the bounded candidates shared by the generators. We retain 150 local pattern types ranked by the number of supporting mining tasks. For global structures and templates, we cycle through the six source collections, taking one candidate from each available collection per round until the limits in the table are reached. Attachment rules connect the retained pattern types. Pattern and attachment probabilities use unit smoothing; interface-coverage reweighting is disabled.

\begin{table}[!t]
\centering
\small
\renewcommand{\arraystretch}{1.12}
\begin{tabularx}{\columnwidth}{@{}>{\raggedright\arraybackslash}Xrr@{}}
\toprule
{\bfseries Component} & {\bfseries Extracted} & {\bfseries Shared candidates} \\
\midrule
\multicolumn{3}{@{}l}{\textit{Local structures}} \\
Subtask types & 14 & 14 \\
Local patterns & 300 & 150 \\
Attachment rules & 7,900 & 194 \\
\midrule
\multicolumn{3}{@{}l}{\textit{Global structures}} \\
Global structures & 100 & 24 \\
Refinement rules & 288 & 60 \\
Integration templates & 252 & 24 \\
Coupling templates & 4 & 4 \\
\bottomrule
\end{tabularx}
\caption{Sizes of extracted structure libraries and shared candidates for controlled generation.}
\label{tab:learned-library}
\end{table}

\paragraph{Targets and generators.}
Each condition has 60 requests per scale, with seeds 41, 53, and 67 contributing twenty requests each. Smaller graphs require 7--11 terminal groups and 2--4 integration groups; larger graphs require 14--18 and 4--6. Integration counts non-delivery groups receiving contributions from at least two distinct upstream groups. The source maximum is eleven terminal groups, so fixed-global generation has no eligible larger-scale source.

Programmatic composition uses at most six global operations, six local attempts, and 24 attachment steps, allowing local expansion to exceed a group's target by at most two leaf subtasks. Direct LLM generation uses DeepSeek-Flash at temperature 0.5, a 32,768-token completion limit, and one call per request without repair. It receives the shared library, schema, targets, resource criteria, and two mining-side format demonstrations. The demonstrations specify the output format; one falls below the resource-supported work minima, which are stated explicitly in the request.

\paragraph{Comparisons and sensitivity.}
The target-deviation weights are 1 for group count and 2 for integration count. Candidate penalties add 2 for reusing a source task and 1 for reusing a source collection; a refinement into at least three children without an integration group adds 0.75 when its collection has already been used. A refinement with no internal bindings adds 0.25. Reuse is measured over previously applied templates in the same composition.

The unguided variant removes target-directed selection and stopping while retaining the local routine. Operation-cap analysis uses 0, 2, 4, 6, and 8 operations on paired initializations and local random streams. Intervals use 2,000 request-bootstrap draws within seed blocks. Outputs at different caps are paired by request. Wilson intervals summarize proportions. Failed submissions remain in the denominator, including six transport failures and one truncated direct-generation response.

\begin{table*}[!t]
\centering
{\small
\setlength{\tabcolsep}{5pt}
\renewcommand{\arraystretch}{1.16}
\begin{tabularx}{\textwidth}{@{}>{\raggedright\arraybackslash}p{.07\textwidth}>{\raggedright\arraybackslash}p{.17\textwidth}>{\raggedright\arraybackslash}X@{}}
\toprule
\multicolumn{3}{@{}p{\textwidth}@{}}{\textbf{Task.} Restore a Python build frontend. Its CLI must invoke the build engine and isolated environment to produce build artifacts.} \\
\midrule
\textbf{Time} & \textbf{Actor / event} & \textbf{Selected trajectory excerpt} \\
\midrule
4:15 & Developer / edit & Patches \texttt{\_builder.py}, including backend validation and exception handling, then runs component tests. \\
4:18 & Tool / result & \texttt{env.py:59: NotImplementedError: Implement capability: DefaultIsolatedEnv.\_\_init\_\_} \\
\addlinespace[4pt]
\rowcolor[HTML]{DCEDC1}
4:37 & Planner $\rightarrow$ developer & ``I (planner) am taking \texttt{src/build/\_\_main\_\_.py} --- do NOT edit it. You own \texttt{src/build/env.py} and \texttt{src/build/\_builder.py}. Focus on env.py first'' \\
\addlinespace[4pt]
5:03 & Planner / edit & Writes the CLI, including \texttt{from .env import DefaultIsolatedEnv} and \texttt{with DefaultIsolatedEnv(installer=installer) as env:}. \\
5:06 & Tool / local test & \texttt{3 passed in 0.03s} (CLI helper and module tests). \\
\addlinespace[4pt]
5:25 & Tool / broader test & \texttt{FAILED tests/test\_main.py::test\_build\_isolated[False] - NotImplementedError} \newline
\texttt{29 failed, 14 passed, 1 warning in 1.38s} \\
\addlinespace[4pt]
5:39 & Planner / repair & Changes the default output-directory handling in the CLI and reruns argument-parsing tests. \\
5:42 & Tool / local retest & \texttt{13 passed, 1 warning in 0.06s} (argument-parsing tests). \\
\midrule
\rowcolor[HTML]{A8E6CF}
\multicolumn{3}{@{}p{\textwidth}@{}}{\textbf{Final evaluation.} Project, hooks, and artifacts pass; isolation and CLI fail. \textbf{3/5 goals completed; task unresolved.}} \\
\bottomrule
\end{tabularx}
\caption{Chronological excerpts from a Meta-Team build-tool episode. Times are elapsed minutes and seconds. Message, code, and tool-output fragments reproduce the record; edit descriptions summarize operations. The final row reports complete task grading after the episode.}
\label{tab:trajectory-cases}
}
\end{table*}

\paragraph{Structural acceptance criteria.}
\label{app:resource-audit}
All four criteria use schema and interface checks, acyclic dependencies, and the same terminal-group range. Let $N$ denote the terminal-group count. Common requirements include at least $2N-1$ leaf subtasks, at least $\lceil(N-1)/2\rceil$ groups with multiple leaves, an explicit coupling, a resource shared across groups, and capability and delivery groups.

\textit{Initial.} Count integration at non-delivery terminal groups with incoming bindings from at least two distinct upstream groups. This count must lie in the requested integration interval.

\textit{Resource-backed.} Count a subtask toward the work minimum only when it participates in a compatible explicit artifact dependency or shared-resource access. A shared resource requires at least two participating nodes and a writer; cross-group artifacts require a matching typed boundary binding. At least $2N-1$ subtasks and $\lceil(N-1)/2\rceil$ groups with two such subtasks must qualify. Integration is counted from artifacts supplied by distinct upstream groups, using the same requested interval and excluding delivery groups.

\textit{With delivery.} Apply the resource-backed criterion but include delivery groups in the integration count. The upper and lower integration bounds remain unchanged, so an added delivery group can push an output above the upper bound.

\textit{All subtasks.} Apply the resource-backed criterion and additionally require every leaf subtask to qualify. Figure~\ref{fig:resource-audit-sensitivity} compares all four criteria on the same generated outputs.

\begin{table*}[!t]
\centering\small
\setlength{\tabcolsep}{5pt}\renewcommand{\arraystretch}{1.16}
\begin{tabularx}{\textwidth}{@{}>{\raggedright\arraybackslash}p{.07\textwidth}>{\raggedright\arraybackslash}p{.17\textwidth}>{\raggedright\arraybackslash}X@{}}
\toprule
\multicolumn{3}{@{}p{\textwidth}@{}}{\textbf{Task.} Prepare a radio programme and corrected transcripts. Each transcript must retain the speaker label, attribution, and embargo specified for its segment.} \\
\midrule
\textbf{Time} & \textbf{Actor / event} & \textbf{Selected trajectory excerpt} \\
\midrule
\rowcolor[HTML]{DCEDC1}
10:54 & Coordinator $\rightarrow$ worker-4 & Delegates the rights register: ``Compute the correct rights manifest per task rules''. \\
\addlinespace[4pt]
13:22 & Worker-4 / write & Creates \texttt{NCR RESULT rights}. For SEG001: \texttt{speaker\_label: Contributor 01}; \texttt{attribution: Northbank source SEG001}; \texttt{embargo: 2027-04-19}. \\
\addlinespace[4pt]
13:24 & Tool / receipt & The rights-note creation succeeds. \\
\addlinespace[4pt]
\rowcolor[HTML]{DCEDC1}
37:12 & Coordinator $\rightarrow$ worker-14 & Delegates transcripts with fields ``\texttt{speaker\_label}: exact allowed label'' and ``\texttt{attribution}: exact source attribution''; requests that both new notes be read back. \\
\addlinespace[4pt]
40:16 & Worker-14 / write & Creates \texttt{NCR RESULT transcripts}. For SEG001: \texttt{speaker\_label: Presenter}; \texttt{attribution: Riverbend Voices archive}; \texttt{embargo: 2027-04-19}. \\
\addlinespace[4pt]
40:16 & Tool / receipt & The transcript-note creation succeeds. \\
\addlinespace[4pt]
\midrule\rowcolor[HTML]{A8E6CF}
\multicolumn{3}{@{}p{\textwidth}@{}}{\textbf{Final relation check.} For SEG001, the embargo agrees, but the speaker label and attribution differ between registers. Both artifacts exist; the required transfer is incomplete.} \\
\bottomrule\end{tabularx}
\caption{Chronological excerpts from an AOrchestra radio episode. Times are elapsed minutes and seconds. Short quotations reproduce instruction fragments; operation rows show selected fields. The relation check compares final artifacts, separately from tool-call success.}\label{tab:case-aorchestra-rights}\end{table*}

\section{Case Study}
\label{app:case-study}
\label{app:collaboration-cases}
The relation statistics in RQ2 assess whether final outputs satisfy task dependencies and shared-resource constraints. The cases below trace how Meta-Team, AOrchestra, Magentic-One, and Superagent divide work, pass plans or results between agents, and respond to feedback during execution.
\medskip

\paragraph{An upstream dependency remains incomplete.}
In the build-frontend task, the CLI depends on an isolated environment to execute builds (Table~\ref{tab:trajectory-cases}). Meta-Team assigns the CLI to its planner and the environment to its developer. The developer's test exposes an unimplemented environment constructor, but the planner subsequently writes a CLI that calls it. Local CLI tests pass, while broader tests expose the unresolved dependency. A later argument-parsing repair improves local test results without completing the environment. Final evaluation reports failures in isolation and CLI, with the latter also affected by a separate backend-import failure.

\paragraph{Consistency across outputs.}
The volunteer case in the main text provides a complementary example. In its 15M-token Meta-Team episode, worker-b handles bookings and worker-a handles calendar participation. Corrections reduce mismatched dates from eleven to six, leaving 14 of 20 volunteer records consistent. Both outputs exist, but the required agreement is incomplete. By comparison, a conservation episode has matching calendar and staffing records across sixteen sessions created by separate workers under a shared coordinator. This episode provides a contrasting case in which separately produced outputs satisfy their shared scheduling requirement.

\begin{table*}[!t]
\centering\small
\setlength{\tabcolsep}{5pt}\renewcommand{\arraystretch}{1.16}
\begin{tabularx}{\textwidth}{@{}>{\raggedright\arraybackslash}p{.07\textwidth}>{\raggedright\arraybackslash}p{.17\textwidth}>{\raggedright\arraybackslash}X@{}}
\toprule
\multicolumn{3}{@{}p{\textwidth}@{}}{\textbf{Task.} Restore a Python build frontend with a working CLI, build engine, and isolated environment. The delivered components must support independent consumer execution.} \\
\midrule
\textbf{Time} & \textbf{Actor / event} & \textbf{Selected trajectory excerpt} \\
\midrule
\rowcolor[HTML]{DCEDC1}
12:13 & Orchestrator $\rightarrow$ Coder & ``Implement \texttt{src/build/\_\_main\_\_.py} in full'' and ``Emit each as one executable sh block for ComputerTerminal.'' \\
\addlinespace[4pt]
12:25 & Coder / proposal & ``I'll implement \texttt{\_\_main\_\_.py} in full''; returns a shell block writing the CLI and running tests. \\
\addlinespace[4pt]
12:28 & Orchestrator $\rightarrow$ terminal & ``Execute Coder's latest fenced sh block exactly as written'' and report stdout/stderr. \\
\addlinespace[4pt]
12:30 & Tool / test & \texttt{ERROR collecting tests/test\_main.py}: the \texttt{flaky} marker is unregistered. Tests stop during collection. \\
\addlinespace[4pt]
13:16 & Orchestrator $\rightarrow$ terminal & Relays a later Coder block registering the marker and rerunning the utility and CLI tests. \\
\addlinespace[4pt]
13:17 & Tool / retest & \texttt{registered flaky marker}; then \texttt{AttributeError: module 'build.\_\_main\_\_' has no attribute '\_NO\_COLORS'}. \\
\addlinespace[4pt]
\rowcolor[HTML]{DCEDC1}
13:22 & Orchestrator $\rightarrow$ Coder & ``find every module-level attribute it references on \texttt{build.\_\_main\_\_} (starting with \texttt{\_NO\_COLORS}) and implement them''. \\
\addlinespace[4pt]
\midrule\rowcolor[HTML]{A8E6CF}
\multicolumn{3}{@{}p{\textwidth}@{}}{\textbf{Final evaluation.} Project, hooks, artifacts, and isolation pass; CLI fails. \textbf{4/5 goals completed; task unresolved.}} \\
\bottomrule\end{tabularx}
\caption{Chronological excerpts from a Magentic-One build-tool episode. The Coder proposes code; ComputerTerminal executes it through the orchestrator. Message fragments reproduce the record, while descriptions summarize operations. Local test errors and final goal checks are reported separately.}\label{tab:case-magentic-build}\end{table*}

\begin{table*}[!t]
\centering\small
\setlength{\tabcolsep}{5pt}\renewcommand{\arraystretch}{1.16}
\begin{tabularx}{\textwidth}{@{}>{\raggedright\arraybackslash}p{.07\textwidth}>{\raggedright\arraybackslash}p{.17\textwidth}>{\raggedright\arraybackslash}X@{}}
\toprule
\multicolumn{3}{@{}p{\textwidth}@{}}{\textbf{Task.} Relocate a museum exhibition after its planned venue becomes unavailable. Arrange exhibit inspections, installation, school visits, and public opening with qualified staff and non-overlapping shifts for each employee.} \\
\midrule
\textbf{Time} & \textbf{Actor / event} & \textbf{Selected trajectory excerpt} \\
\midrule
\rowcolor[HTML]{DCEDC1}
16:46 & Worker $\rightarrow$ subagent & Delegates a 219-entry work plan: ``Each row must become ONE native submitted Shift Assignment''. States: ``There are no duplicate rows requiring dedupe EXCEPT if you are resuming''. \\
\addlinespace[4pt]
16:55--17:08 & Subagent / read & Reads the shared plan. Six separate rows, including inspections of exhibits OBJ001 and OBJ005, assign employee E002 to the same June 29 shift, 08:00--17:00. \\
\addlinespace[4pt]
17:31 & Tool / receipt & The subagent's first assignment for E002 on June 29 is submitted successfully. \\
\addlinespace[4pt]
17:51 & Subagent / create & Requests another assignment for E002 on June 29 using the same shift. The application returns a conflict. \\
\addlinespace[4pt]
18:00--18:08 & Subagent / submit & Submitting further shifts for E002 on dates already covered by submitted shifts returns three conflicts. The subagent inspects the saved records. \\
\addlinespace[4pt]
18:12 & Episode / stop & The shared token budget is exhausted. Nine new shift records exist, five submitted; the required staffing summary is missing. \\
\midrule\rowcolor[HTML]{A8E6CF}
\multicolumn{3}{@{}p{\textwidth}@{}}{\textbf{Final evaluation.} Venue arrangements, programme planning, and consent and funding records pass; transport, staffing, financial reconciliation, and handover fail. \textbf{3/7 goals completed; task unresolved.}} \\
\bottomrule\end{tabularx}
\caption{Superagent's plan-to-execution handoff in an exhibition-recovery task. Times are elapsed minutes and seconds. Quotations reproduce delegation instructions; other rows summarize the plan, tool receipts, and final state.}
\label{tab:case-superagent-exhibition}
\end{table*}

\setcounter{dbltopnumber}{2}
\begin{table*}[t]
\begin{mypromptbox}[Root Task Abstraction]{promptGreen}
{\rmfamily\bfseries Task Definition:} Convert the task specification into a structured description of its required outcomes and resource interfaces.\par\smallskip
{\rmfamily\bfseries Input:} Task specification: \{task\_specification\}; source materials: \{source\_materials\}.\par\smallskip
{\rmfamily\bfseries Requirements:}
\begin{itemize}[noitemsep, topsep=0pt, parsep=0pt, leftmargin=*]
\item Identify the overall objective and the artifacts or information needed to accomplish it.
\item Distinguish required inputs, produced outputs, and access to mutable state.
\item Define observable acceptance criteria supported by the source materials.
\end{itemize}
{\rmfamily\bfseries Output Format:}\par
\{goal, consumes, produces, acceptance, accesses\}\par
\end{mypromptbox}
\par\vspace{3pt}
\caption{Prompt template for extracting task objectives, inputs, outputs, and resource accesses.}
\label{fig:abstraction-prompts}
\end{table*}

\paragraph{AOrchestra: transfer between delivered artifacts.}
Table~\ref{tab:case-aorchestra-rights} follows rights-register and transcript preparation by different workers. Both notes are created successfully. For SEG001, the transcript retains the embargo date but replaces the speaker label and source attribution. The final artifacts thus expose an incomplete dependency even though both deliverables are present.

\paragraph{Magentic-One: execution feedback across specialized roles.}
Table~\ref{tab:case-magentic-build} follows Coder's implementation through ComputerTerminal execution and feedback to the orchestrator. The orchestrator first addresses test-collection settings, then asks Coder to implement missing CLI attributes. The team uses the reported errors to direct repairs, but the CLI remains incomplete at termination. Both software excerpts use the same task and 5M-token limit.

\paragraph{Superagent: shared-resource conflicts in a delegated plan.}
The exhibition task (Project Management) requires qualified staff for exhibit inspections, installation, school visits, and public opening. Each employee's shifts must not overlap. The worker prepares a staffing plan and delegates creation of the corresponding application records to a subagent (Table~\ref{tab:case-superagent-exhibition}). It requests one shift assignment per plan row, although six rows assign the same employee to the same date and 08:00--17:00 shift. The subagent submits the first assignment, then encounters conflicts when creating or submitting repeated assignments. At the token limit, five new shifts are submitted, but the required staffing summary linking activities to shifts is missing. The handoff passes on conflicting staff assignments that remain unresolved during execution.

\section{Prompts}
\label{app:prompts}
Tables~\ref{fig:abstraction-prompts} and~\ref{fig:recursive-prompt} present task abstraction and recursive decomposition; Table~\ref{fig:grounding-prompts} presents scenario refinement and outcome requirement extraction. Table~\ref{fig:analysis-prompts} contains the preliminary analysis template and the Direct LLM baseline for controlled graph composition. The templates summarize the instructions and output fields used at each stage. The Direct LLM configuration is detailed in Appendix~\ref{app:controlled-generation-details}. In its output schema, \texttt{roles} assigns subtask types, \texttt{partition} maps subtasks to groups, \texttt{values} records produced and consumed artifacts, and \texttt{states} records shared-resource accesses.

\begin{table*}[t]
\begin{mypromptbox}[Recursive Task Decomposition]{promptBlue}
{\rmfamily\bfseries Task Definition:} Decompose the parent task into independently assignable subtasks with distinct local outcomes.\par\smallskip
{\rmfamily\bfseries Input:} Parent task: \{parent\_task\}; supporting evidence: \{source\_materials\}.\par\smallskip
{\rmfamily\bfseries Requirements:}
\begin{itemize}[noitemsep, topsep=0pt, parsep=0pt, leftmargin=*]
\item Each child must have a coherent objective, explicit interfaces, and checkable acceptance criteria.
\item Child outputs must jointly cover the parent objective and preserve its external interfaces.
\item Reference existing resources consistently and identify the owner of each produced output.
\item Keep routine implementation steps within a subtask. If no meaningful decomposition exists, return an empty child list.
\end{itemize}
{\rmfamily\bfseries Output Format:}\par
\{"children": [\par
  \{goal, consumes, produces, acceptance, accesses\}, ...\par
]\}\par
\end{mypromptbox}
\par\vspace{3pt}
\caption{Prompt template for recursive task decomposition.}
\label{fig:recursive-prompt}
\end{table*}

\begin{table*}[t]
\begin{mypromptbox}[Scenario Refinement]{promptGreen}
{\rmfamily\bfseries Task Definition:} Refine a task graph into a concrete scenario supported by the available environment.\par\smallskip
{\rmfamily\bfseries Input:} Composed task graph: \{task\_graph\}; environment capabilities: \{environment\_capabilities\}; scenario context: \{scenario\_context\}.\par\smallskip
{\rmfamily\bfseries Requirements:}
\begin{itemize}[noitemsep, topsep=0pt, parsep=0pt, leftmargin=*]
\item Specify substantive work branches and their local outcomes.
\item Adapt the work to available entities, artifacts, and permitted operations.
\item Describe how outputs connect work in the resulting scenario.
\item Identify shared resources and consistency requirements between branches.
\item State initial conditions, constraints, and verifiable completion requirements.
\end{itemize}
{\rmfamily\bfseries Output Format:}\par
Background and objective; facts and constraints; work branches; dependencies and shared resources; completion conditions.\par
\end{mypromptbox}
\medskip
\begin{mypromptbox}[Outcome Requirement Extraction]{promptBlue}
{\rmfamily\bfseries Task Definition:} Extract verifiable acceptance requirements from the public task description.\par\smallskip
{\rmfamily\bfseries Input:} Numbered task evidence: \{task\_description\}.\par\smallskip
{\rmfamily\bfseries Requirements:}
\begin{itemize}[noitemsep, topsep=0pt, parsep=0pt, leftmargin=*]
\item Associate each requirement with supporting task evidence.
\item Preserve quantities, dates, identity constraints, exceptions, and permitted choices.
\item Distinguish required changes from facts that must remain unchanged.
\item Group related requirements into outcome goals; accept alternative solutions that satisfy them.
\item Record unresolved ambiguities without inventing additional obligations.
\end{itemize}
{\rmfamily\bfseries Output Format:}\par
\{"clauses": [...], "goals": [...], "global\_clause\_ids": [...], "ambiguities": [...]\}\par
\end{mypromptbox}
\par\vspace{3pt}
\caption{Prompt templates for scenario refinement and extraction of outcome requirements.}
\label{fig:grounding-prompts}
\label{fig:verifier-prompt}
\end{table*}

\begin{table*}[t]
\begin{mypromptbox}[Task Structure Analysis]{promptGreen}
{\rmfamily\bfseries Task Definition:} Analyze the work structure supported by the task materials.\par\smallskip
{\rmfamily\bfseries Input:} Public task: \{task\_description\}; available initial materials: \{task\_materials\}.\par\smallskip
{\rmfamily\bfseries Requirements:}
\begin{itemize}[noitemsep, topsep=0pt, parsep=0pt, leftmargin=*]
\item Identify substantial branches with distinct local objectives and checkable results.
\item Distinguish separately progressable branches from sequential phases of one deliverable.
\item Identify consequential output dependencies and shared-state interactions; common inputs alone do not establish either.
\item Support judgments with cited evidence. Do not impose collaboration structure when the materials do not support it.
\item Retain unknown judgments where the available evidence is insufficient.
\end{itemize}
{\rmfamily\bfseries Output Format:}\par
Branches and supporting evidence; dependency and coupling relations; progressability and interdependence judgments; remaining uncertainty.\par
\end{mypromptbox}
\medskip
\begin{mypromptbox}[Direct LLM Generation (Construction Baseline)]{promptBlue}
{\rmfamily\bfseries Task Definition:} Generate a task graph in one response using the supplied task structure library and structural targets.\par\smallskip
{\rmfamily\bfseries Input:} Task structure library: \{task\_structure\_library\}; demonstrations: \{examples\}; structural targets: \{targets\}.\par\smallskip
{\rmfamily\bfseries Requirements:}
\begin{itemize}[noitemsep, topsep=0pt, parsep=0pt, leftmargin=*]
\item Construct the task hierarchy and the corresponding leaf-level dependency graph.
\item Satisfy the specified group-count, integration, and common structural requirements.
\item Use compatible subtask types and input/output and shared-resource interfaces.
\item Represent artifact handoffs through producer--consumer relationships and coupling through compatible shared-state bindings.
\item Ensure consistent references, complete interfaces, and acyclic directed dependencies.
\end{itemize}
{\rmfamily\bfseries Output Format:}\par
\{tree, bindings, roles, partition, dep, cpl, values, states, external\_inputs\}\par
\end{mypromptbox}
\par\vspace{3pt}
\caption{Prompt templates for preliminary task-structure analysis and the Direct LLM baseline in the graph-composition experiment.}
\label{fig:preliminary-prompt}
\label{fig:analysis-prompts}
\end{table*}

\section{Extended Related Work}
\label{sec:related}

\subsection{LLM-based Multi-Agent Systems}
LLM-based MAS organize work through communication and specialized roles. CAMEL studies role-playing interactions, AutoGen provides configurable conversation patterns, and MetaGPT and ChatDev coordinate software design, implementation, and testing~\citep{li2023camel,wu2023autogen,hong2023metagpt,qian2024chatdev}.

Automatic system design includes optimizing prompts and graph connections in GPTSwarm, generating communication topologies in G-Designer, and sampling query-dependent architectures in MaAS \citep{zhuge2024gptswarm,zhang2024gdesigner,zhang2025maas,feng2026graph}. AFlow searches code-represented workflows using execution feedback \citep{zhang2024aflow}. Recent work explores dynamic sub-agent creation, decentralized coordination, structural adaptation, and evolutionary design \citep{ruan2026aorchestra,yang2025agentnet,liu2026masfly,hu2026evomas,cao2026evoteam,xiang2026systematic}. G-Memory and Meta-Team reuse collaboration experience and study team-level self-evolution \citep{zhang2025gmemory,hao2026metateam}. MAC-Bench evaluates such systems through tasks with interconnected work requirements.

\subsection{General-Purpose Agent Benchmarks}
General-purpose benchmarks assess agents through completed actions and outcomes in interactive environments. AgentBench spans multiple domains, while GAIA emphasizes reasoning, information gathering, and tool use~\citep{liu2024agentbench,mialon2024gaia}. SWE-bench, Terminal-Bench, and TravelPlanner focus on repository changes, terminal work, and constrained planning~\citep{jimenez2024swebench,merrill2026terminalbench,xie2024travelplanner}. WebArena and OSWorld extend evaluation to web and desktop interaction; AppWorld couples interactive coding with application-state verification~\citep{zhou2024webarena,xie2024osworld,trivedi2024appworld}.

WorkArena++ constructs compositional knowledge-work tasks, and TheAgentCompany evaluates workplace activities involving websites, coding, and simulated coworkers \citep{boisvert2024workarenaplus,xu2024agentcompany}. $\tau$-bench evaluates tool use and policy compliance during user interaction; $\tau^2$-bench adds dual control of a shared environment \citep{yao2024taubench,barres2025tau2bench}. These settings extend evaluation to composed activities and interaction with users or coworkers.

\subsection{Benchmarks for Multi-Agent Collaboration}
Dedicated collaboration benchmarks vary the information and interaction conditions facing a team. COMMA uses complementary multimodal observations, SILO-BENCH distributes information across agents, and AgentsNet varies communication networks~\citep{ossowski2024comma,zhang2026silobench,grotschla2025agentsnet}. MASBench and PAC-Bench examine collaboration under partial observability and privacy constraints~\citep{chu2026masbench,park2026pacbench}. ProtocolBench and MAS-Orchestra study communication protocols and orchestration through controlled tasks~\citep{du2026protocolbench,ke2026masorchestra}.

MultiAgentBench provides milestone-based evaluation of collaborative and competitive scenarios \citep{zhu2025multiagentbench}. MAgIC and Collab-Overcooked examine social reasoning and cooperation in games \citep{xu2024magic,sun2025collabovercooked}. TeamBench enforces planner, executor, and verifier roles \citep{kim2026teambench}; ALEM and AgentWorld study open-ended coordination and long-horizon collaboration \citep{tessera2026alem,shu2026agentworld}. MAC-Bench composes executable tasks with varied collaboration requirements while leaving agent roles and communication topologies open.

Analyses of coordination, failures, and agent scaling identify the importance of communication, specification, verification, and task-dependent organization \citep{agashe2025llmcoordination,cemri2025mast,kim2025agentscaling}. MASEval evaluates complete agentic systems, including orchestration and execution infrastructure \citep{emde2026maseval}. These studies motivate complementing task outcomes with process and resource measurements.

\subsection{Task Structure and Environment Generation}
AgentGen generates environments from domain text and evolves tasks within them \citep{hu2024agentgen}. EnvScaler and ScaleEnv synthesize tool-interactive environments and tasks with programmatic validation \citep{song2026envscaler,tu2026scaleenv}. SWE-smith modifies repositories to produce failures against existing tests \citep{yang2025swesmith}. MedAgentGym provides code-centric biomedical tasks, while Gym-Anything converts software into agent environments \citep{xu2026medagentgym,aggarwal2026gymanything}. Compositional construction also appears in WorkArena++ and the atomic task generator of $\tau^2$-bench \citep{boisvert2024workarenaplus,barres2025tau2bench}.

Structured representations support planning, coordination, and retrieval. Hierarchical task-network planning represents complex activities through decomposition \citep{erol1994umcp}; coordination theory and joint-action models describe dependencies and commitments \citep{malone1994coordination,cohen1991teamwork,grosz1996sharedplans,tambe1997flexibleteamwork}. KG-Translator translates queries into graph-level clues for passage retrieval \citep{liuretrieval}. Network motifs characterize recurring local interaction structures \citep{milo2002networkmotifs}. MAC-Bench mines reusable local patterns and global structures from selected tasks, composes them, and instantiates executable evaluation tasks.